\documentclass[twocolumn,english,aps,prl,showpacs,superscriptaddress]{revtex4-1}
\usepackage[latin9]{inputenc}
\usepackage{amsmath}
\usepackage{amssymb}
\usepackage{graphicx}
\usepackage{esint}
\usepackage{physics}
\usepackage{enumitem}
\usepackage{braket}
\usepackage{spverbatim}
\makeatletter

\usepackage{physics}
\usepackage{xcolor}
\usepackage{multirow}
\makeatother

\usepackage{babel}

\begin{document}

\title{Bulk-corner correspondence of time-reversal symmetric insulators: \\ 
deduplicating real-space invariants 
}
\author{Sander Kooi}
\affiliation{Institute for Theoretical Physics, Center for Extreme Matter and
Emergent Phenomena,\\ Utrecht University, Princetonplein 5, 3584
CC Utrecht, the Netherlands}

\author{Guido van Miert}
\affiliation{Dipartimento di Fisica ``E. R. Caianiello'',
Universit\`a di Salerno I-84084 Fisciano (Salerno), Italy }

\author{Carmine Ortix}
\affiliation{Institute for Theoretical Physics, Center for Extreme Matter and
Emergent Phenomena,\\ Utrecht University, Princetonplein 5, 3584
CC Utrecht, the Netherlands}
\affiliation{Dipartimento di Fisica ``E. R. Caianiello'',
Universit\`a di Salerno I-84084 Fisciano (Salerno), Italy }

\date{\today}

\newcommand{\CO}[1]{{\color{blue}#1}}
\newcommand{\Com}[1]{{\color{black}#1}}
\newcommand{\GM}[1]{{\color{red}#1}}
\newcommand{\SK}[1]{{\color{violet}#1}}

\begin{abstract}
The topology of insulators is usually revealed through the presence of gapless boundary modes: this is the so-called bulk-boundary correspondence. However, the many-body wavefunction of a crystalline insulator is endowed with additional topological properties that do not yield surface spectral features, but manifest themselves as (fractional) quantized electronic charges localized at the crystal boundaries. Here, we formulate such bulk-corner correspondence for the physical relevant case of materials with time-reversal symmetry and spin-orbit coupling. To so do we develop ``partial" real-space invariants that can be neither expressed in terms of Berry phases nor using symmetry-based indicators. These new crystalline invariants govern the (fractional) quantized corner charges both of isolated material structures and of heterostructures without gapless interface modes.  We also show that the partial real-space invariants are able to detect all time-reversal symmetric topological phases of the recently discovered fragile type. 
\end{abstract}

\maketitle

\section{Introduction}

The discovery of topological insulators has fundamentally challenged our common classification of materials in terms of electrical insulators and electrical conductors \cite{has10,qi11}. Topological insulators are in fact materials that are insulating in their bulk but allow for perfect conduction of electrical currents along their surfaces. 
This macroscopic physical property is the immediate consequence of the  topological properties of the  ground state of the insulator: this is the essence of the so-called bulk-boundary correspondence. 
In a topological insulator, the electrical conduction is protected against different detrimental effects since the surface electronic modes have an anomalous nature. 
The chiral states appearing at the edge of a quantum Hall insulator, for instance,  represent anomalous states since in a conventional one-dimensional atomic chain it is impossible to find a different number of left-moving and right-moving electrons \cite{tho82,hal82}. The helical edge states of quantum spin-Hall insulators \cite{kan05b,kan05,fu06,ber06,mol07}, as well as the single Dirac surface states of strong three-dimensional topological insulators \cite{fu07topological,moo07,fu07b,zha09,ras13,xia09} violating the fermion doubling theorem, are other prime examples of such anomalies.

When unitary spatial symmetries are taken into account, additional topological crystalline phases can arise \cite{fu11,hsi12,hsi14,liu14b,ses16}. The non-trivial topology of the system then guarantees the presence of anomalous surface states appearing only on surfaces that are left invariant under the ``protecting" crystalline symmetry, and which violate stronger versions of the fermion doubling theorem~\cite{fan17}. Furthermore, crystalline symmetries can lead to a class of insulating phases, dubbed higher-order topological insulators, with conventionally gapped surface states but with anomalous gapless states appearing on the hinges connecting two surfaces related to each other by the crystalline protecting symmetry \cite{sch18,sch18b,mie18,kha18,koo18,kha18c}.

The single Slater determinant describing the ground state of a non-interacting crystalline insulator generally possesses additional topological indices that are not immediately related to the presence or absence of anomalous gapless surface states. For instance, the 
electric polarization of an inversion-symmetric one-dimensional atomic chain is either integer or semi-integer, with a quantized value that does not depend upon microscopic details, but is rather encoded in a gauge-invariant topological index \cite{van93}.
More recently, it has been shown that 
 excess electronic charges localized at various topological defects, such as dislocations, can be (fractionally) quantized, thus representing yet other incarnations of bulk quantities encoded in topological invariants \cite{lau16,mie17,mie18r,li20}. Quantized charges appearing at the corners and disclinations of two-dimensional crystals have been very recently measured in metamaterials \cite{liu20,pet20b,pet20} and proposed to appear in recently synthesized materials structures \cite{pha20e}.
Together with the topological indices dictating the presence of anomalous gapless surface modes, the gauge-invariant bulk quantities governing the appearance of quantized defect charges specify 
the entire  ``observable" topological content of a crystal. 

In systems with broken time-reversal symmetry this  set of crystalline topological invariants can be entirely expressed in terms of the symmetry properties of the occupied single particle Bloch states at the high-symmetry points of the Brillouin zone 
with the addition of the  Chern number. However, for the physically relevant case of materials with spin-orbit coupling and time-reversal symmetry eigenvalues-based schemes do suffer of intrinsic limitations. Topological crystalline phases with robust boundary modes may pass completely undetected~\cite{kooi19b} using the current classification schemes based on symmetry indicators \cite{bra17,po17,zha19,ver19,tan19}. Likewise, the real-space invariants originally introduced in Ref.~\cite{mie18} are insufficient to determine the quantized excess charges. This is because Kramers' theorem inevitably doubles the electronic charges, making the real-space invariants partially, often completely, trivial. 
Progress can be made identifying (partial) Berry phase~\cite{lau16} invariants and/or using Wilson loops as topological indices \cite{koo19,sch19,bra19} as exemplified by the bulk-disclocation charge correspondence of rotation-symmetric two-dimensional crystals~\cite{mie18r}. This additional knowledge, however, does not completely determine the (fractional) quantized electronic charges at the crystal boundaries. 

Here, we overcome these hurdles by developing a novel strategy that for  two-dimensional time-reversal symmetric insulators in rotation symmetric crystals is able to fully resolve 
this missing
bulk-corner 
correspondence. The crux of our analysis is the ability to effectively deduplicate the real-space invariants of Ref.~\cite{mie18}, 
using a new computationally efficient framework that requires only the knowledge of the Bloch wavefunctions throughout the Brillouin zone. We show that the resulting partial real-space invariants 
govern not only the quantized corner charges of insulators that are deformable to atomic limit, but also determine the quantized corner charges in heterostructures comprising topologically distinct quantum spin-Hall insulators. 
Even more importantly, the bulk-corner correspondence we formulate here allows the detection of all topological states of the fragile type \cite{po18,son19} in time-reversal symmetric crystals, in much the same way as the recently introduced twisted bulk-boundary correspondence is able to diagnose fragile phases in systems without spin-orbit coupling \cite{son20}.

\section{Corner charges are topological bulk invariants}
We first discuss the intrinsic limitations of symmetry-based eigenvalue schemes in detecting the (fractionally) quantized corner charges of time-reversal symmetric insulators. 
Let us consider for simplicity a crystal with a simple twofold rotation symmetry ${\mathcal C}_2$. 
At the high-symmetry points in the Brillouin zone (BZ)  $\Gamma = (0,0)$, $X = (\pi,0)$  , $Y = (0,\pi)$, and $M = (\pi,\pi)$,  the ${\mathcal C}_2$ symmetry provides us with eight natural numbers $\Gamma_{\pm i},\ldots,M_{\pm i}$, which denote  the multiplicities of occupied Bloch states with rotation eigenvalues $\pm i$ (from here onwards we will consider systems of spin $1/2$ fermions). 
These multiplicities, taken by themselves, define proper integer invariants since they can only change by bandgap closing and reopening processes. However, the multiplicities at different momenta are not linearly independent because of the presence of the compatibility relations $HS_i+HS_{-i}=N_F$ with $HS=\Gamma,X,Y,M$ and $N_F$ the number of occupied bands. Even more importantly, the rotation symmetry multiplicities do not correspond to any known physical observable. 

\begin{figure}
\includegraphics[width=1\columnwidth]{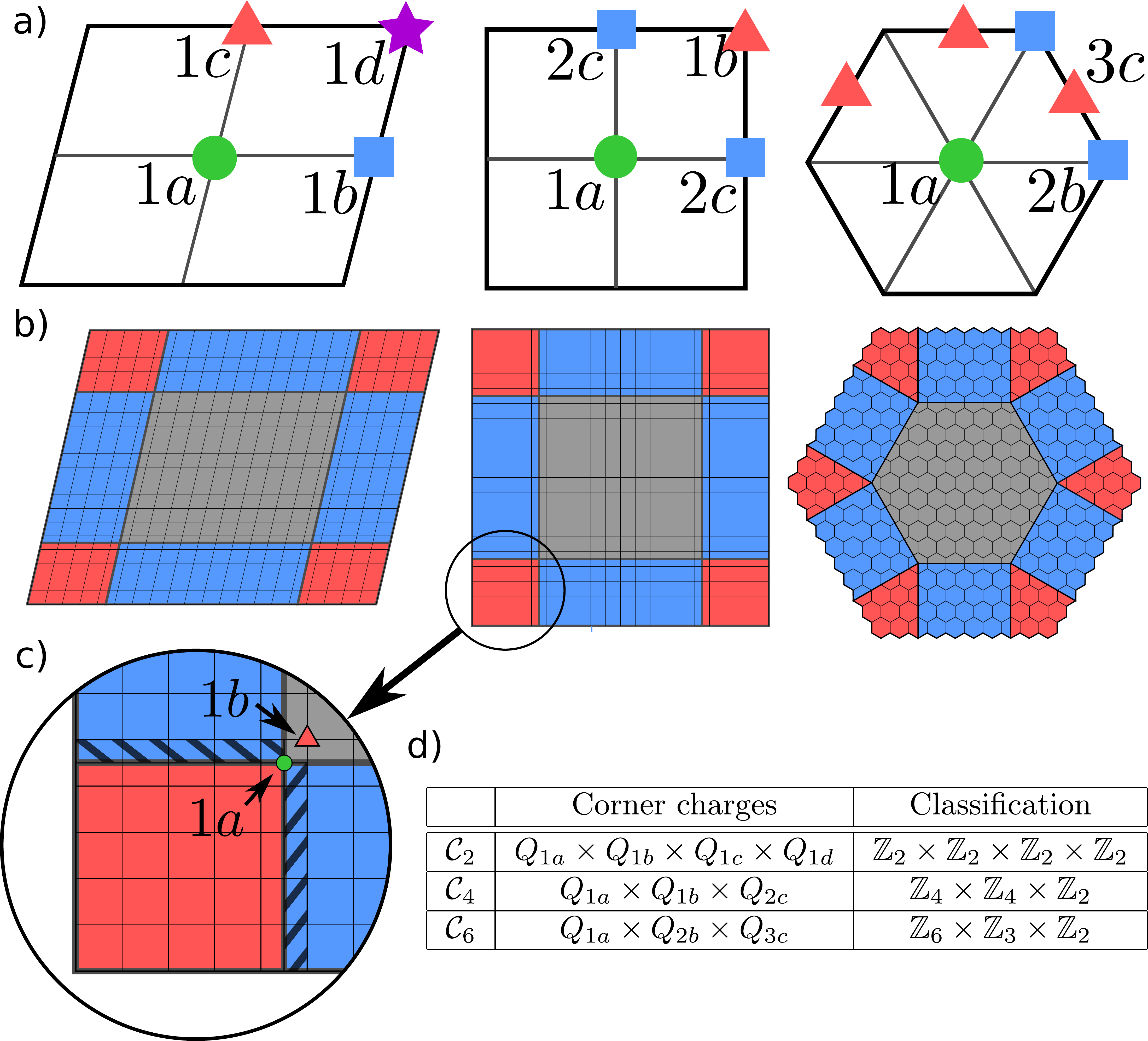}
\caption{(a) Unit cells of rotation-symmetric crystals. The special Wyckoff positions and their multiplicities are explicitly indicated. (b) Rotation-symmetric crystals in open-disk geometries. The red, blue and grey regions indicates bulk, edges, and corners respectively. Note that when considering a reference ${\mathcal C}_2$-symmetric Wyckoff position, the corner charge has to be summed over the adjacent corners not related by ${\mathcal C}_2$ symmetry in order to represent a quantized bulk quantity. 
(c) Zoom in of the ${\mathcal C}_4$-symmetric corners with the two possible reference Wyckoff positions. (d) Table of the corner charge topological characterization for
$\mathcal{C}_{2},\mathcal{C}_{4}$ and $\mathcal{C}_{6}$ symmetric
systems. \label{fig:1}}
\end{figure}

These shortcomings can, however, be overcome by constructing the linear combinations of the multiplicities originally introduced in Ref.~\onlinecite{mie18}, recently dubbed real-space invariants~\cite{son20}. 
For a $\mathcal{C}_2$-symmetric insulator this approach gives rise to four $\mathbb{Z}$-numbers $\nu_{1a}, \ldots, \nu_{1d}$ in one-to-one correspondence with the four $\mathcal{C}_2$-symmetric Wyckoff positions $1a,\ldots,1d$ [c.f. Fig.~\ref{fig:1}].  As a result, we find a global ${\mathbb Z}^4$ classification, which is fully in agreement with $K$-theory studies~\cite{kru17b}. 
Moreover, the parities of these real-space invariants dictate the values of the fractional part of the quantized charge contained in corners measured with respect to ${\mathcal C}_2$-symmetric Wyckoff positions [c.f. Fig.~\ref{fig:1}]. For example, if $\nu_{1a}$ is an even (odd) integer then the corner charge $Q_{1a}$ measured with respect Wyckoff position $1a$ is equal to $0$ ($1/2$) $\textrm{ mod }1$, {\it i.e.} 
$Q_{1a}=\nu_{1a}/2 \textrm{ mod }1$. 
Note that only the fractional part of the corner charge represents a proper bulk value, as the possible occurrence of in-gap corner modes affects the integer part. In other words, we cannot distinguish between $Q_{1a}=0$ and $Q_{1a}=1$ from a topological point of view. 

 The above one-to-one correspondence between symmetry labels (encoded in $\nu_{1x}$'s) and the fractional part of the quantized corner charges is completely general and applies to all rotation-symmetric two-dimensional crystals. Specifically, the corner charges measured with respect to a special Wyckoff position with a site symmetry group containing an $n$-fold rotation symmetry has a fractional part quantized in multiples of $1/n$, thereby defining a ${\mathbb Z}_n$ topological invariant. Beside the ${\mathbb Z}_2^4$ classification of ${\mathcal C}_2$-symmetric crystals discussed above, this leads to  ${\mathbb Z}_4 \times {\mathbb Z}_4 \times {\mathbb Z}_2$ classification in fourfold rotation symmetric crystals and a $\mathbb{Z}_6 \times \mathbb{Z}_3 \times \mathbb{Z}_2$ classification [c.f. Fig.~\ref{fig:1}(a),(d)] with the invariants all formulated in terms of symmetry labels [c.f. Appendix A].

Even though the discussion above is 
informative from a purely theoretical point of view and of relevance to metamaterial structures, it has a limited value for the large number of insulating  materials which possess time-reversal symmetry. This is because Kramers' theorem dictates that the corner charges measured with respect to a ${\mathcal C}_n$-symmetric Wyckoff position must be quantized in multiples of $2/n$. This clearly trivializes the ${\mathbb Z}_2^4$ topological content  of twofold rotation symmetric crystals, whereas it leaves two residual $\mathbb{Z}_2$ topological invariants -- corresponding to the (semi)integer values of the corner charges relative to the Wyckoff positions $1a$ and $1b$ [see Fig.~\ref{fig:1}] -- in fourfold rotation symmetric crystals and a $\mathbb{Z}_3 \times \mathbb{Z}_3$ classification in ${\mathcal C}_6$-symmetric crystals. The doubling does not alter the classification of  ${\mathcal C}_3$-symmetric corners, thus implying that in threefold rotation-symmetric crystals real-space invariants can be redefined even in time-reversal symmetric conditions. 

For evenfold rotation-symmetric corners, and this is key, the trivialization of the corner charges is instead only apparent. 
Kramers' theorem not only engenders the doubling of the corner charge quantum from $1/n$ to $2/n$; it further guarantees that microscopic details at the edge and corners of a finite size crystal can only change 
the value of the corner charge by an even integer rather than an integer. This can be immediately seen from the fact that when created by local perturbations in-gap corner modes have to come in pairs. In other words, now the cases $Q_{1a}=0$ and $Q_{1a}=1$ are topologically distinct. 

The fact that in time-reversal symmetric conditions the corner charges {\it modulo} $2$ are  bulk quantities has a twofold effect. First, it implies that from a corner-charge perspective the topological crystalline characterization of rotation-symmetric insulators is the one tabled in Fig.~\ref{fig:1}(d) even in the presence  of time-reversal symmetry. Second, and most importantly, the quantized corner charges cannot be simply expressed in terms of the symmetry eigenvalues: consider the simple case of a ${\mathcal C}_2$-symmetric insulator. Time-reversal symmetry requires  that all the integers $\Gamma_{\pm i} \ldots M_{\pm i} \equiv N_F / 2$ rendering the real-space invariants [c.f. Appendix A and Ref.~\onlinecite{mie18}] completely trivial.

\section{Partial real-space invariants} 

Having established that in time-reversal symmetric conditions the quantized values of the corner charges cannot be entirely read off from the point-group symmetries eigenvalues, we next 
derive the bulk-corner correspondence by formulating crystalline topological invariants entirely different in nature. 
Our approach can be decomposed into three steps: we start from a particularly stringent set of assumptions that will be partially relaxed in each consecutive step by employing the ${\mathcal U}(N_F)$ gauge degree of freedom of the many-body wavefunction. The end product of this endeavour will be a formulation of crystalline invariants that can readily be computed using standard numerical methods.

Let us first notice that as long as we consider insulators whose ground state can be described in terms of exponentially localized Wannier functions, and thus adiabatically connected to an atomic insulator, 
the formulation of the topological invariants governing the quantized corner charges with time-reversal symmetry only requires a bulk expression for the number of Wannier Kramers' pairs centered at the special Wyckoff positions in the unit cell. 
Such a formulation can be immediately achieved by considering a simple subclass of time-reversal invariant insulators, {\it i.e.} systems without sizable spin-orbit coupling. In this materials class, we can
naturally split the space of occupied Bloch states into two sectors 
related to each other by time-reversal symmetry:
sector $I$ 
for spin-up electrons, and sector $II$  for
 the spin-down electrons. 
Importantly, choosing the spin quantization axis perpendicular to the crystalline plane
 each sector enjoys the ${\mathcal C}_n$ rotation symmetry of the lattice. 
 This also implies that the fractional part of the corner charge in each sector can be related to the real-space invariants introduced in Ref.~\onlinecite{mie18} via $Q_{N x}^{I,II} = m \nu^{I}_{m x} / n$ where $m$ indicates the multiplicity of the Wyckoff position $m x$ with respect to which the corner charge is measured. Moreover, time-reversal symmetry guarantees that the corner charges associated to two channels are equal, {\it i.e.} $Q^{I}_{1x} = Q^{II}_{1x}$. As a result, the quantized corner charge of time-reversal symmetric insulators are given by $Q_{m x} = 2~Q_{m x}^{I} = 2~Q_{m x}^{II}= 2~m~\nu^{I}_{m x} / n~\textrm{mod}~2$, and are entirely expressed in terms of the sector invariants $\nu_{m x}^{I,II}$ listed in Appendix A.  
 We naturally dub these integers partial real-space invariants in analogy with the partial Berry phase.
 Although the partial real-space invariants have been derived in the context of ``Wannierazible" insulators, they apply equally well to topological states of the fragile type. These recently discovered topological states 
 cannot be represented in terms of symmetric Wannier functions~\cite{po18,son19}, but at the same they do not feature gapless edge states. Being insulating both in their bulk and along their edges, they are characterized by quantized corner charges at ${\mathcal C}_n$-symmetric corners. Furthermore, the hallmark of fragile phases is their decay into an atomic insulating phase by a proper addition of topologically trivial bands. The additivity of the corner charges under band additions then engenders the validity of the partial real-space invariants.  
 
 The absence of spin-orbit coupling considered so far provides us with a natural splitting of the Bloch states in two sectors related to each other by time-reversal symmetry and separately ${\mathcal C}_n$-symmetric. However, and this is key, taking advantage of the ${\mathcal U}(N_F)$ gauge degree of freedom that leaves the Slater determinant unchanged, such a decomposition can be always achieved, even in systems with a sizable spin-orbit coupling. The problem of determining the corner charge therefore boils down to finding a continuous, periodic and rotation-symmetric gauge for two time-reversed sectors, and subsequently computing the real-space invariants for a single sector. There is, however, a small caveat. Namely, in our derivation we have implicitly been assuming that the Chern number per sector vanishes. This follows from the fact that the real-space invariants from Ref.~\onlinecite{mie18} only apply to systems without chiral edge states. Hence, by relying on precisely those invariants we have to demand that the Chern numbers of each sector vanish, i.e. $C^{I} \equiv C^{II} \equiv 0$.  Note that this represents an additional constraint on the gauge as time reversal symmetry only guarantees that $C^{I} + C^{II} \equiv 0$.
 
 \begin{table}
\begin{tabular}{|c|l|}
\hline 
$\mathcal{C}_{2}$ & {$\begin{aligned}
\nu_{1a}^{I} & =\frac{1}{2}C ^{I}+\Gamma_{i}^{I}+\frac{1}{2}\left[\Gamma_{-i}^{I}-X_{-i}^{I}-Y_{-i}^{I}-M_{-i}^{I}\right]\\
\nu_{1b}^{I} & =\frac{1}{2} C^{I}-\frac{1}{2}\left[\Gamma_{-i}^{I}-X_{-i}^{I}+Y_{-i}^{I}-M_{-i}^{I}\right]\\
\nu_{1c}^{I} & =\frac{1}{2} C^{I}-\frac{1}{2}\left[\Gamma_{-i}^{I}+X_{-i}^{I}-Y_{-i}^{I}-M_{-i}^{I}\right]\\
\nu_{1d}^{I} & =\frac{1}{2} C^{I}-\frac{1}{2}\left[\Gamma_{-i}^{I}-X_{-i}^{I}-Y_{-i}^{I}+M_{-i}^{I}\right]
\end{aligned}$}
\tabularnewline
\hline 
$\mathcal{C}_{4}$ & 
{$\begin{aligned}
\nu_{1a}^{I} & =-\frac{1}{2}C^{I}+\left(-3\Gamma_{e^{i\pi/4}}^{I}-\frac{3}{2}\Gamma_{e^{i3\pi/4}}^{I}-\Gamma_{e^{-i3\pi/4}}^{I}-\frac{3}{2}\right.\\
 & \left.\times\Gamma_{e^{-i\pi/4}}^{I}+\frac{3}{2}M_{e^{i3\pi/4}}^{I}+2M_{e^{-i3\pi/4}}^{I}+\frac{3}{2}M_{e^{-i\pi/4}}^{I}+X_{-i}^{I}\right)\\
\nu_{1b}^{I} & =-\frac{1}{2}C^{I}+\left(\frac{3}{2}\Gamma_{e^{i3\pi/4}}^{I}+2\Gamma_{e^{-i3\pi/4}}^{I}+\frac{3}{2}\Gamma_{e^{-i\pi/4}}^{I}\right.\\
 & \left.-\frac{1}{2}M_{e^{i3\pi/4}}^{I}-2M_{e^{-i3\pi/4}}^{I}-\frac{1}{2}M_{e^{-i\pi/4}}^{I}-X_{-i}^{I}\right)\\
\nu_{2c}^{I} & =\frac{1}{2}C^{I}+\frac{1}{2}\left(\Gamma_{e^{i3\pi/4}}^{I}+\Gamma_{e^{-i\pi/4}}^{I}-M_{e^{i3\pi/4}}^{I}-M_{e^{-i\pi/4}}^{I}\right)
\end{aligned}$} 
\tabularnewline
\hline 
$\mathcal{C}_{6}$ & 
{$\begin{aligned}
\nu_{1a}^{I} & =-\frac{1}{2}C^{I}+\left(-5\Gamma_{e^{i\pi/6}}^{I}-\frac{5}{2}\Gamma_{e^{i\pi/2}}^{I}-\Gamma_{e^{i5\pi/6}}^{I}-\frac{1}{2}\Gamma_{e^{-i5\pi/6}}^{I}\right.\\
 & \left.-\Gamma_{e^{-i\pi/2}}^{I}-\frac{5}{2}\Gamma_{e^{-i\pi/6}}^{I}+\frac{3}{2}M_{-i}^{I}+2K_{-1}^{I}+2K_{e^{-i\pi/3}}^{I}\right),\\
\nu_{2b}^{I} & = C^{I}+\left(\Gamma_{e^{i\pi/2}}^{I}+\Gamma_{e^{i5\pi/6}}^{I}+\Gamma_{e^{-i\pi/2}}^{I}+\Gamma_{e^{-i\pi/6}}^{I} \right. \\ 
& \left. -K_{-1}^{I}-K_{e^{-i\pi/3}}^{I}\right),\\
\nu_{3c}^{I} & =\frac{1}{2}C^{I}+\frac{1}{2}\left(\Gamma_{e^{i\pi/2}}^{I}+\Gamma_{e^{-i5\pi/6}}^{I}+\Gamma_{e^{-i\pi/6}}^{I}-M_{-i}^{I}\right).
\end{aligned}$}
\tabularnewline
\hline 
\end{tabular}
\caption{Expressions for the partial real-space invariants 
in $\mathcal{C}_{2}$, $\mathcal{C}_{4}$ and $\mathcal{C}_{6}$
symmetric systems. These formulas are independent of the channel Chern number, but require a continuous periodic, and symmetric set of projectors.}
\label{tab:realspaceinv}
\end{table}

Our second step consists in relaxing precisely this additional constraint on the sector Chern numbers $C^{I,II}$. 
To do so, we will introduce new topological integers that reduce to the expressions of the crystalline invariants of Appendix A and  Ref.~\onlinecite{mie18} whenever the ground state of the insulator is decomposed in two time-reversed channels with zero Chern number. 
 We discuss the rationale behind the definition of these modified topological invariants by discussing a paradigmatic microscopic model.
Consider a bilayer system consisting of two Kane-Mele models~\cite{kan05b} on a uniaxially strained honeycomb lattice, with the two models differing only by the relative sign of the intrinsic spin-orbit coupling strength $\lambda_{ISO}$. Being the sum of two quantum spin-Hall insulators, the bilayer system does not possess metallic edge states, and therefore the bulk corner correspondence is well posed. 
We first determine the partial real-space invariants by decomposing the space of occupied Bloch states according to their spin $s_z$ eigenvalues. 
Having chosen the sign of $\lambda_{ISO}$ opposite in the two layers, each of the spin state has a vanishing Chern number. Therefore, we can safely use the formulation of the  invariants in terms of the individual channel symmetry eigenvalues. It can be easily shown that the non-zero multiplicities of the residual ${\mathcal C}_2$ rotation symmetry are $\Gamma^I_{i} = X^I_i = Y^I_i = M^I_{-i} =2$. Consequently, the partial real-space invariant [c.f. Appendix A]  
$$\nu_{1a}^I = \Gamma^I_{i}+\frac{1}{2}[\Gamma^I_{-i} - X^I_{-i}-Y^I_{-i}-M^I_{-i}] = 1$$
immediately predicts a quantized corner charge measured with respect to the center of the unit cell $Q_{1a} = \nu_{1a}^I~\textrm{mod}~2= 1~\textrm{mod}~2$. Next, we employ a different decomposition wherein a channel  is composed by a spin state, say spin up, in the first layer and the opposite spin state, spin down, in the second layer. This leads to different ${\mathcal C}_2$ multiplicities $\Gamma_{\pm i} = X_{\pm i} = Y_{\pm i} = M_{\pm i} = 1$ accompanied by a non-vanishing Chern number in the two sectors $C^{I,II} = \pm 2$, which clearly violates the bulk corner correspondences since $\nu^{I}_{1x} \equiv 0$. However, if we redefine the partial real-space invariant by accounting for an explicit Chern number contribution, {\it i.e.} 
\begin{equation}
\nu^I_{1a} =\frac{1}{2}C^I+\Gamma^I_{i}+\frac{1}{2}[\Gamma^I_{-i} - X^I_{-i}-Y^I_{-i}-M^I_{-i}],
\label{eq:realspaceinv}
\end{equation}
the ${\mathcal C}_2$-protected topological indices, modulo $2$, are independent of the specific channel decomposition: for the specific case of the Kane-Mele bilayer model $\nu^{I}_{1a}~\textrm{mod}~2 \equiv 1$. This is  sufficient to resolve the quantized bulk corner charges of a twofold rotation symmetric insulator in time-reversal conditions. 
Note that these redefined partial real-space invariants [c.f. Table.~\ref{tab:realspaceinv} for their expressions also in the ${\mathcal C}_{4,6}$ cases]  cannot be applied, {\it per se}, in systems with an odd channel Chern number. These systems realize quantum spin-Hall insulators and, in isolation, do not have a well-defined bulk-corner correspondence because of their gapless edges.

\begin{figure}
\includegraphics[width=1\columnwidth]{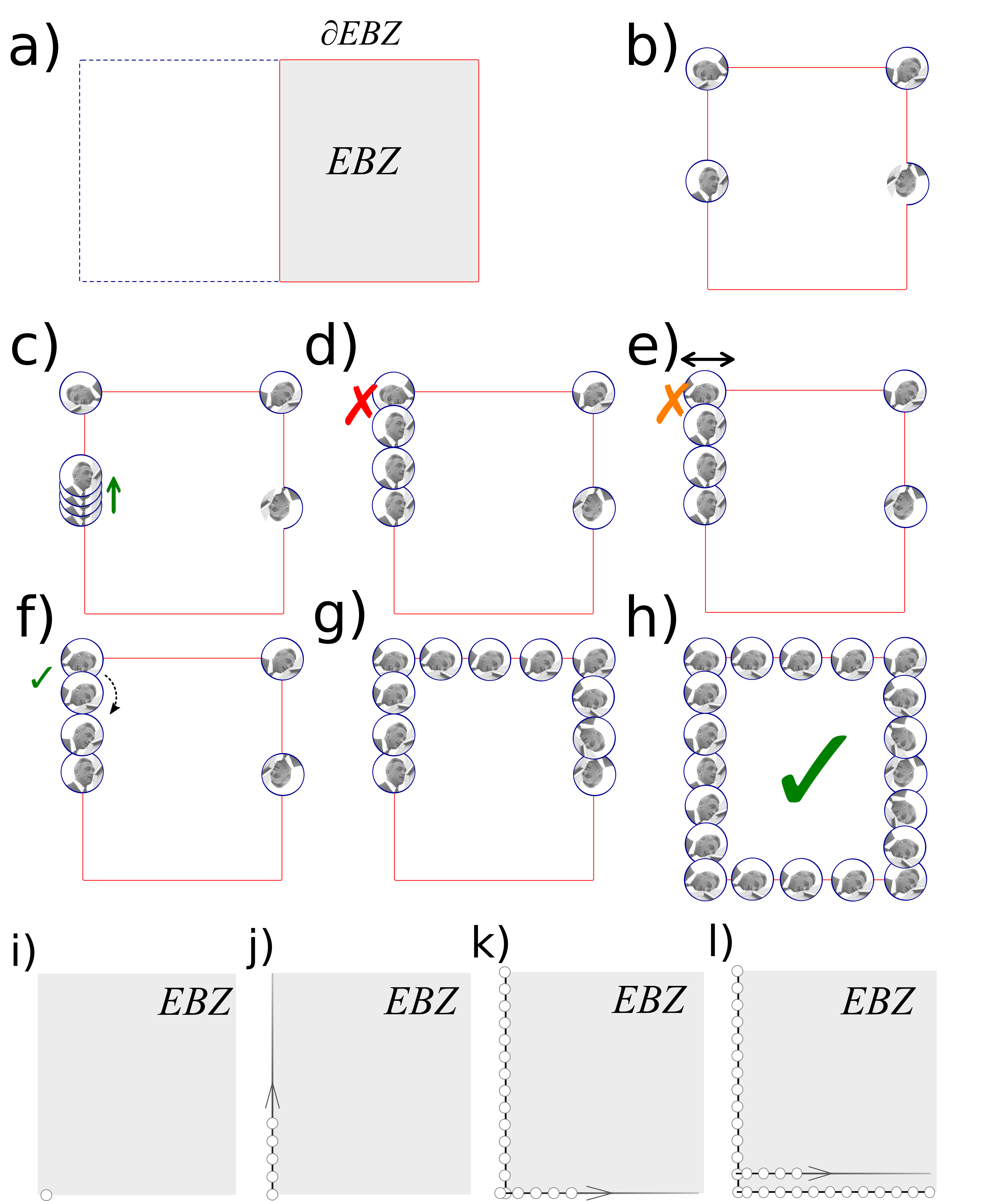}
\caption{Schematic overview of the procedure to construct a $\mathcal{C}_2$ and $\Theta$-symmetric gauge along the effective Brillouin zone boundary.  The gauge at each point is represented by a portrait of Felix Bloch. The $\mathcal{C}_2\Theta$-respecting gauge transformations are rotations of Bloch's portrait, possibly followed by a reflection.  (a) Brillouin zone, with in grey the effective Brillouin zone. (b) selection of $\mathcal{C}_2$- and $\Theta$-symmetric target Bloch states at the four high-symmetry points within EBZ. (c) A continuous gauge is obtained along the line connecting $\Gamma$ and $Y$ by parallel transporting the target state at $\Gamma$ in the direction of $Y$. (d) Topological mismatch between the parallel-transported states at $Y$ and the target Bloch states at $Y$. (e) The topological mismatch at $Y$ reduces to a trivial mismatch upon redefining the target states at $Y$. The redefinition is represented as a mirror operation on the portrait of Bloch. (f) The trivial mismatch is removed by a rotation that is evenly spread out over the parallel-transported states along the line connecting $Y$ and $\Gamma$. (g) Pictorial representation of continuous $\mathcal{C}_2$ and $\Theta$-symmetric gauge along the upper half of the effective Brillouin zone boundary. (h) Extension of the continuous $\mathcal{C}_2$- and $\Theta$-symmetric gauge along the entire effective Brillouin zone boundary. (i)-(l) Procedure to construct a $\mathcal{C}_2$- and $\Theta$-symmetric gauge in the interior of the EBZ by parallel transport.  \label{fig:1Dproc}}
\end{figure}

Having at hand explicit expressions without constraints on the channel Chern number immediately implies that the partial real-space invariants can be computed if we are provided with a continuous and periodic set of projectors $\rho^{I,II}(q) = \sum_m \ket{\Psi^{I,II}_m(\vec{q})}  \bra{\Psi^{I,II}_m(\vec{q})}$, related to each other by time-reversal symmetry and individually rotation symmetric. This is different from the former construction of a set of smooth, periodic and symmetric  Bloch waves $\ket{\Psi_m^{I,II}(q)}$ throughout the entire Brillouin zone, which necessitates individual ``Wannierazible" channels. More importantly,  we can now employ our third step and relax the constraint on the continuity, periodicity and symmetry requirements on $\rho^{I,II}(q)$. As before, let 
us consider for simplicity 
${\mathcal C}_2$ crystals and assume to have hypothetically found a set of continuous, periodic and rotation symmetric projectors $\rho^{I,II}(q)$. 
First, we observe that this is precisely equivalent to having a continuous, smooth and periodic set of Bloch waves $\ket{\Psi^{I,II}_m(\vec{q})}$ in the
 effective Brillouin zone $EBZ = [0,\pi]\times [-\pi,\pi]$ (see Fig.\ref{fig:1Dproc}) 
 such that
\begin{enumerate}[label=(\roman*)]
\item the sewing matrix $S_{\mathcal{C}_2}(\vec{q}) = \langle\Psi^{\alpha}_m(-\vec{q})|\mathcal{C}_2|\Psi^{\beta}_n(\vec{q})\rangle $ is block-diagonal along the 
two rotation symmetric high-symmetry lines $q_x = 0, \pi$;
\item the sewing matrix $S_{\mathcal{C}_2\Theta}(\vec{q}) = \langle\Psi^{\alpha}_m(\vec{q})|\mathcal{C}_2\Theta|\Psi^{\beta}_n(\vec{q})\rangle $ is 
block off-diagonal
in the entire
EBZ. In particular the Bloch waves in each channel can be redefined  to satisfy $\ket{\Psi^{II}_m(\vec{q})} = \mathcal{C}_2\Theta \ket{\Psi^{I}_m(\vec{q})}$, in which case the sewing matrix  $S_{\mathcal{C}_2\Theta}(\vec{q})^{\alpha,\beta}_{m,n} = \sigma_x^{\alpha,\beta}\delta_{m,n}$ with $\sigma_x$ the first Pauli matrix and $\delta_{m,n}$ the Kronecker delta.
\end{enumerate}
We next use that since the Bloch waves along the boundary ${\partial EBZ}$ of the EBZ are smooth, periodic and satisfy the symmetry constraints, the Chern number contribution to the partial real-space invariants can be rewritten using Stokes' theorem as the contour integral of the Berry connection 
\begin{align}
\dfrac{1}{2} C^{I} \equiv \frac{1}{2 \pi}\oint_{\partial EBZ} \mathrm{d}\vec{q}\cdot Tr(\vec{A}^I(\vec{q})),
\label{eq:chern1}
\end{align}
where we emphasize that the equation above is a true equality, and does not have any integer ambiguity. 
Considering that the multiplicities of the rotation symmetry eigenvalues are also uniquely determined by the Bloch waves along $\partial EBZ$, one could conclude that the computation of the partial real-space invariants, {\it e.g.} Eq.~\ref{eq:realspaceinv}, can be reduced to an effective one-dimensional problem. 

However, Eq.~\eqref{eq:chern1} constitutes a true equality only if the Bloch waves in both time-reversed channels are continuous, periodic and symmetric throughout the entire effective Brillouin zone. If we were to abandon this constraint, then the halved Chern number $C^I/2$ would be only determined modulo an integer that is insufficient to determine, modulo $2$, the partial real-space invariants. 
To make further progress, 
let us consider the effect upon the smooth Bloch waves in the entire EBZ of a  ${\mathcal U}(N_F)$ transformation that preserves the block off-diagonal form of the ${\mathcal C}_2 \Theta$ sewing matrix. We will refer to this subset of unitary transformations as ${\mathcal U}_{{\mathcal C}_2 \Theta}(N_F)$. It can be shown   that this group of transformations constitutes a subgroup conjugate to the orthogonal group ${\mathcal O}(N_F)$ [see Appendix B and Ref.~\footnote{E.P. van den Ban, private communication}]. As a result, we  have that the homotopy classes of ${\mathcal U}_{{\mathcal C}_2 \Theta}(N_F)$ are the same as those of the orthogonal group. Specifically,
${\mathcal U}(N_F)_{\mathcal{C}_2\Theta}$ consists of two connected components:
\begin{align*}
\pi_0({\mathcal U}(N_F)_{\mathcal{C}_2\Theta})=\mathbb{Z}_2, 
\end{align*}
corresponding to the subset of matrices with determinant equal to $+1$ and $-1$.
Moreover, as the first homotopy group of the orthogonal group is non-trivial, we find that the same applies to ${\mathcal U}(N_F)_{\mathcal{C}_2\Theta}$. In particular, this yields:
\begin{align*}
\pi_1({\mathcal U}(N_F)_{\mathcal{C}_2\Theta})) =   
\begin{cases} \mathbb{Z} \textrm{ if } N_F = 2,\\
\mathbb{Z}_2 \textrm{ if } N_F > 2.
\end{cases} 
\end{align*}
In other words, a map from a closed loop to 
our group of unitary transformations
is either characterized by a $\mathbb{Z}$- or by a $\mathbb{Z}_2$-winding number. 
Since 
the effective Brillouin zone boundary, $\partial EBZ$, defines a closed loop, 
we can introduce the $\mathbb{Z}_2$-type winding number $W({\mathcal U}|_{\partial EBZ})$.
This ${\mathbb Z}_2$ winding number is of utmost importance since it allows us to uniquely determine how the halved sector Chern number, modulo $2$, transforms under a ${\mathcal U}_{{\mathcal C}_2 \Theta}$ transformation.  
We can therefore introduce new partial real-space invariants that are identical, modulo $2$, to the expressions of Table~\ref{tab:realspaceinv} and thus correctly capture the bulk-corner correspondence. For instance, the analog of Eq.~\ref{eq:realspaceinv} can be written as 
\begin{align}
\nu^I_{1a} &= \frac{1}{2\pi}\oint_{\partial EBZ} \mathrm{d}\vec{q}\cdot Tr(\vec{A}^I(\vec{q}))+W(U|_{\partial EBZ}) \nonumber \\
& +\Gamma^I_{i}+\frac{1}{2}[\Gamma^I_{-i} - X^I_{-i}-Y^I_{-i}-M^I_{-i}]
\label{eq:compinvariant}
\end{align}
Since the expression above is invariant under arbitrary gauge transformations along the boundary of the effective Brillouin zone that respect $\mathcal{C}_2$- and $\Theta$-symmetry, the rotation-symmetric time-reversed channels of periodic and smooth Bloch waves $\Psi^{I,II}(\vec{q})$ along the boundary of the effective Brillouin zone $\partial EBZ$
can be chosen completely independent of the 
Bloch waves $\chi^{I,II}(\vec{q})$ in the interior of the EBZ, provided they make the ${\mathcal C}_2 \Theta$ sewing matrix completely off-diagonal.
We have therefore
decomposed the task of finding a 
continuous, periodic, and rotation-symmetric set of time-reversed projectors $\rho^{I,II}(q)$
into two simpler, and computationally possible
problems. Namely, the construction of a 
``$\mathcal{C}_2\Theta$-symmetric gauge"
within 
the
$EBZ$, and the construction of a 
``$\mathcal{C}_2$ \& $\Theta$-symmetric gauge''
along 
the boundary
$\partial EBZ$. 
The winding number 
$W({\mathcal U}|_{\partial EBZ})$ effectively reveals
whether or not the gauge along 
$\partial EBZ$
can be matched with the gauge constructed within the EBZ. 
All in all, 
the following steps need to be implemented in order to compute the partial real-space invariants of a twofold rotation-symmetric crystal with time-reversal symmetry: 
\begin{enumerate}[label=(\roman*)]
\item construct a continuous and periodic $\mathcal{C}_2$ \& $\Theta$-symmetric gauge of Bloch waves
$\Psi(\vec{q})$ 
along $\partial EBZ$,
\item compute the partial Berry phase and the multiplicities of the rotation-symmetry eigenvalues
using $\Psi(\vec{q})$,
\item construct a continuous, but not necessarily periodic, $\mathcal{C}_2\Theta$-symmetric gauge for Bloch waves $\chi(\vec{q})$ within the $EBZ$, and
\item compute the $\mathbb{Z}_2$- or $\mathbb{Z}$-winding number of the overlap-matrix $U(\vec{q}) = \langle\chi(\vec{q}|\Psi(\vec{q})\rangle$ along $\partial EBZ$.
\end{enumerate}

Fig.~\ref{fig:1Dproc} sketches a computationally feasible strategy to perform the four steps mentioned above. We refer the reader to the Methods section for more details on the computational procedures in twofold rotation-symmetric crystals and Appendix C for the generalization to the case of  $\mathcal{C}_4$- and $\mathcal{C}_6$-symmetric crystals.

\section{Relation to topological crystalline line invariants}
We next discuss the relation between the partial real-space invariants defining the bulk-corner correspondence and 
the quantized partial polarizations -- the so-called ``line''  invariants -- along lines in the Brillouin zone that are mirror, or equivalently twofold-rotation, symmetric \cite{lau16,mie17}. 
We recall that these  $\mathbb{Z}_2$ invariants can be written in a ${\mathcal U}(N_F)$ invariant form that only requires a periodic gauge from numerically obtained eigenstates \cite{mie17}. 

Let us  consider, as before, the case of twofold rotation symmetric crystal [we refer the reader to Appendix D for the ${\mathcal C}_{4,6}$ cases] and use that insisting on a gauge choice that makes the two time-reversed channels ${\mathcal C}_2$ symmetric the quantized partial polarization along the $k_2=0$ ($\Gamma - X$) line of the Brillouin zone (BZ) can be simply written as $\gamma^I_{1}= \Gamma_i^I + X_i^I$ modulo an even integer. 
Likewise, the quantized partial polarization along the $k_1=0$ ($\Gamma - Y$) line of the BZ reads $\gamma^I_{1}= \Gamma_i^I + Y_i^I$. 
Note that the quantized partial polarization on the $k_{1,2}=\pi$, {\it i.e.} the $Y-M$ and the $X-M$ lines respectively, are not independent since they are related to the polarization $\gamma_{1,2}^{I}$ via the ${\mathbb Z}_2$ topological invariant $\nu_{FKM}$ originally introduced by Fu, Kane and Mele~\cite{kan05,fu06}. It is straightforward to show that the two independent $\mathbb{Z}_2$ line invariants can be immediately expressed in terms of the partial real-space invariants. Using that the latter are well-defined modulo $2$ in a ${\mathcal C}_2$-symmetric crystal, and taking advantage of the compatibility relations for the ${\mathcal C}_2$ symmetry eigenvalues $HS_{i} + HS_{-i} \equiv N_F$ with $HS=\Gamma,X,Y,M$, we immediately find [c.f. Table.~\ref{tab:realspaceinv}] the following equalities modulo 2
\begin{align*}
\left(\nu_{1b}^{I}+\nu_{1d}^{I}\right)  & =\left(\nu_{FKM}-\gamma_{1}^{I}\right), \\
\left(\nu_{1c}^{I}+\nu_{1d}^{I}\right)  & =\left(\nu_{FKM}-\gamma_{2}^{I}\right). 
\end{align*}
Here, we have used the relation between the Chern number of the $I,II$ channels and the Fu-Kane-Mele invariant: $\nu_{FKM} = C^I \mod 2$. 
While partial real-space invariants uniquely determine the topological crystalline line invariants, the opposite is not true. 
More generally, the fact that the line invariants 
do not resolve the bulk-corner correspondence 
 can be seen by noticing that there is a single constraint on the $\nu_{1x}^I$ invariants modulo 2, namely
\begin{align*}
\left(\nu_{1a}^{I}+\nu_{1b}^{I}+\nu_{1c}^{I}+\nu_{1d}^{I}\right) & =\Gamma_{i}^{I}-\Gamma_{-i}^{I} = \dfrac{N_{F}}{2}. 
\end{align*}
This also implies that for an arbitrary $N_F$ number of occupied bands, a ${\mathcal C}_2$- and $\Theta$-symmetric insulator can be characterized by three partial real-space $\mathbb{Z}_2$ indices, which, together with $\nu_{FKM}$, form a $ \mathbb{Z}_2 \times \mathbb{Z}_2 \times \mathbb{Z}_2 \times \mathbb{Z}_2$ classification. This has clearly more topological content than the 
 $ \mathbb{Z}_2 \times \mathbb{Z}_2 \times \mathbb{Z}_2 $ characterization in terms of line invariants and Fu-Kane-Mele invariants. Notice that the same holds true also in crystals 
  with fourfold- and sixfold-rotation symmetries. This is simply because the corresponding ${\mathbb Z}_{4,6}$ quantized corner charges cannot be resolved by the ${\mathbb Z}_2$ quantized polarizations. 
  
  \begin{figure}
\includegraphics[width=1\columnwidth]{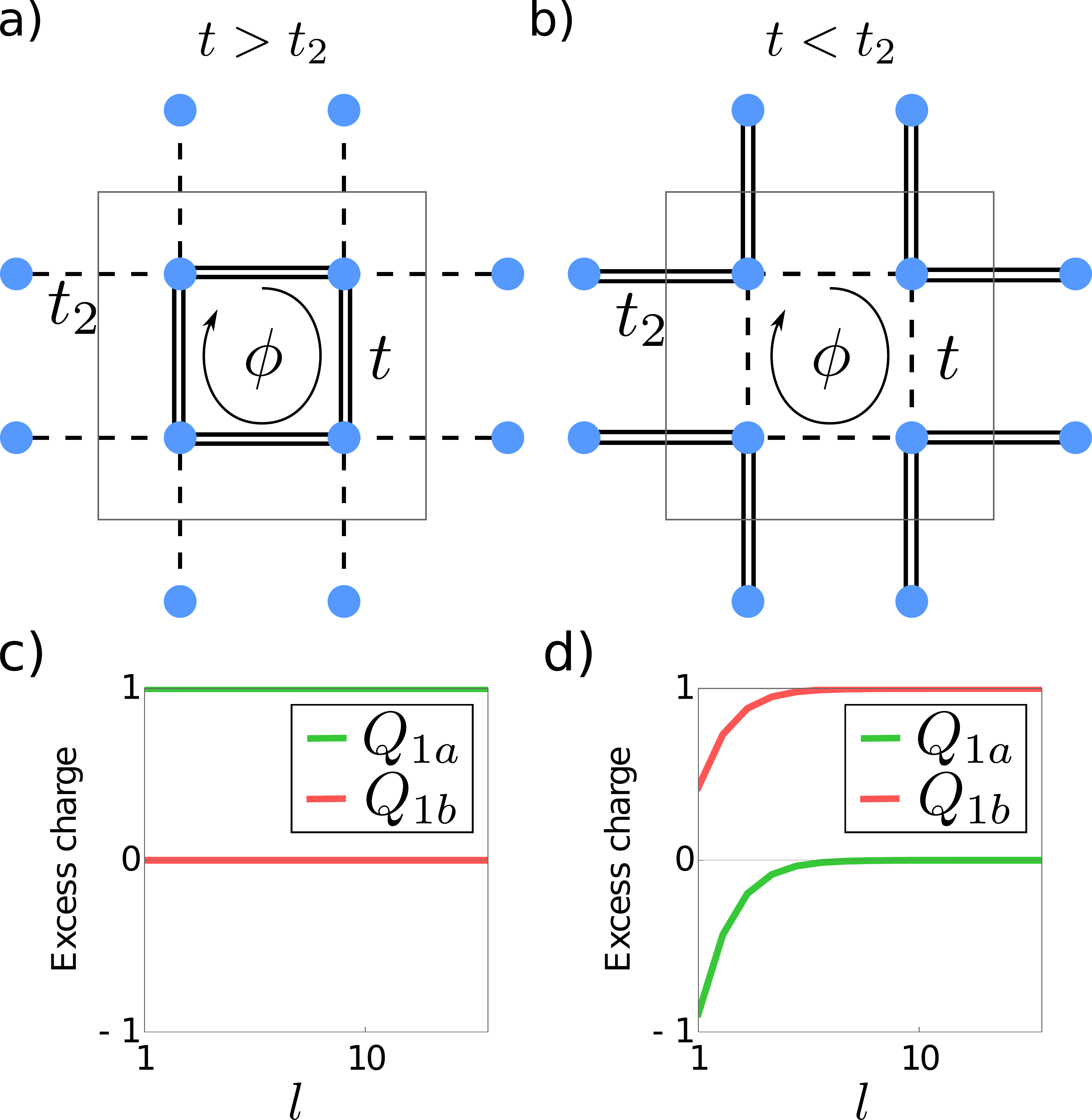}
\caption{(a) Sketch of the Hamiltonian for one spin sector of the $\mathcal{C}_4$-symmetric system that realizes an atomic insulating phase at half-filling with two electrons at Wyckoff position $1a$ or $1b$ (b) depending on the parameters. This model is adapted from Ref.~\cite{ben17}. (c) Behavior of the corner charge density for the phase where two Kramers pairs are localized at $1a$ and (d) $1b$ as obtained by increasing the $l \times l$ corner region size. Note that both phases have the same line invariants and symmetry eigenvalues.
\label{fig:C4line}}
\end{figure}

 To prove concretely that the topological crystalline line invariants do not completely resolve the bulk-corner correspondence let us introduce 
a fourfold rotation-symmetric atomic insulating phase at $N_F=4$ using the model schematically shown in Fig.\ref{fig:C4line}. It can be thought of as being composed of two time-reversed copies of the spinless model introduced in Ref.~\cite{ben17} with the two channels mixed by a spin-orbit coupling term \footnote{We couple site 1 and 2 by a term $\lambda i\sigma_x$ and site 3 and 4 by a term $\lambda i\sigma_y$ within the unit cell, and set $\lambda=0.3$}. At half-filling, the insulating state realized for $|t/t_2|>1$ and $\phi=-\pi$  [see  Fig.~\ref{fig:C4line}(a)]  has a simple decomposition in two channels with zero Chern number and  $\nu_{1a}^I \equiv 1; \nu^I_{1b,2c} \equiv 0$. This is consistent with the explicit computation of the corner charge shown in Fig.~\ref{fig:C4line}(c) that shows $Q_{1a}=1$ while $Q_{1b}=Q_{2c}=0$. 
However, tuning the parameters to $|t/t_2|<1$ and $\phi=\pi$ [see Fig.~\ref{fig:C4line}(b)], this model realizes a different set of real-space invariants $\nu_{1b}^I \equiv 1; \nu^I_{1a,2c} \equiv 0$. And indeed the charge density shown in Fig.~\ref{fig:C4line}(d) predicts $Q_{1b}=1$ while $Q_{1a}=Q_{2c}=0$.
The change in charge density is however not detected by the line invariants that are vanishing in both spaces, {\it i.e.} 
$\gamma^I_{1} \equiv \gamma^I_2 \equiv 0$.  This is because the two insulating phases can be described in terms of two Wannier Kramers pairs centered either at the center $1a$ of the unit cell or at the $1b$ edge of the unit cell, which clearly give the same partial polarizations.

\section{Detecting the crystalline topology of quantum spin-Hall insulators}
\label{sec:qshheometry}
\begin{figure}
\includegraphics[width=1\columnwidth]{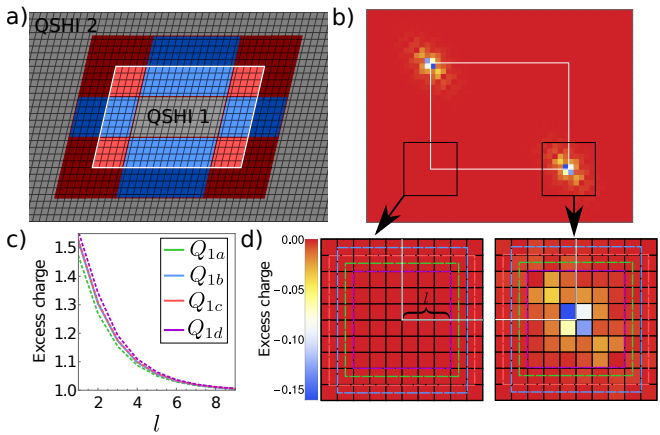}
\caption{(a) Schematic of a $\mathcal{C}_{2}$ QSH system embedded in another QSH system with
periodic boundary conditions. The edge and corner regions of the interface are colored blue and red, respectively. (b) Corresponding charge density for the QSH systems discussed in the text. (c) Behavior of the corner charge as a function of the $l \times l$ corner regions size. (d) Zoom in of the charge density, depicting the two corners and the regions used to calculate the corner charges\label{fig:qshCC}}
\end{figure}

As mentioned above, the partial real-space invariants provide us with the bulk-corner correspondence in crystals that are insulating both in the bulk and along their edges. 
Therefore, they are ill-defined when dealing with quantum spin-Hall insulators due to the presence in the latter of helical edge states. This assertion, however, is only true when considering crystalline systems in isolation. Let us now instead consider one insulating system that is completely
surrounded by 
a second
insulator~\cite{xio20} as shown in Fig.~\ref{fig:qshCC}(a). 
For such a geometry, we 
define
the corner charge of the combined system as the sum of the charge in the corner of the 
inner
insulator and the charge in the 
L-shaped region of the 
surrounding
insulator adjoining the first corner region. 
In addition, both regions will be measured with respect to a special Wyckoff position in order to ensure quantization of the corner charge. In the case of atomic and fragile topological insulators, inspection of the corner charge in this geometry tells us nothing new. 
On the contrary, if the two insulators are of the quantum spin-Hall type, by computing the combined corner charge additional information on the crystalline topology can be extracted.

We recall that from an ``edge" perspective all quantum spin Hall insulators are topologically identical: 
the presence of the helical edge states is mandated by the non-trivial value of the Fu-Kane-Mele $\mathbb{Z}_2$ invariant. In ${\mathcal C}_n$-symmetric crystals, however, quantum spin-Hall phases can be additionally characterized by the quantized partial Berry phases, and they
can be revealed by the charge trapped at dislocation defects. 
As shown in Ref.~\cite{kooi19b}, however, quantum spin-Hall phases in twofold rotation symmetric crystals are endowed with 
an additional topological ${\mathbb Z}_2$ invariant, which cannot be probed at these topological defects. We will now show that the corner charge in the geometry of Fig.~\ref{fig:qshCC}(a) is diagnosed precisely by this crystalline topological index. 

To prove the assertion above, we use  the Kane-Mele model, which, as before, will be considered on a uniaxially strained honeycomb lattice to remove the additional threefold rotation symmetry. 
Using the results of Ref.~\cite{kooi19b}, it can be shown that the quantum spin-Hall insulating phase realized choosing the nearest neighbor hopping amplitude $t=1$ is topologically distinct from the insulating phase realized choosing the opposite sign of the hopping amplitude, even though the spin-orbit coupling $\lambda_{ISO}$ remains unchanged. Moreover, the two states share the same partial Berry phases and consequently cannot be discriminated by analyzing the charge trapped at dislocations. When using the two phases in the combined geometry of Fig.~\ref{fig:qshCC}(a), an explicit computation of the quantized values of the corner charge [see Figs.~\ref{fig:qshCC}(b)-(d)] yields $Q_{1a,1b,1c,1d}=1$, as opposed to the result one would get for a system composed of two quantum spin-Hall insulators with equal crystalline topological indices, {\it i.e.} $Q_{1a,1b,1c,1d}=0$. 

Even more importantly, the quantized corner charges can be straightforwardly obtained using the bulk formulation of the partial real-space invariants, whereas they cannot be read off from the line invariants.  
For a Kane-Mele model on a strained honeycomb lattice with the parameter set choice $(t, \lambda_{ISO})= (1,1)$ the twofold rotation-symmetric channel with Chern number $C^I=1$ is characterized by the following set of rotation symmetry labels $\Gamma^{I}_{i}=X^I_i=Y^I_i=M^I_{-i}=1$.  In the Kane-Mele quantum spin-Hall phase with $(t,\lambda_{ISO})=(-1,1)$ the $I$ channel with $C^I=1$ has reversed rotation symmetry labels $\Gamma^{I}_{-i}=X^I_{-i}=Y^I_{-i}=M^I_{i}=1$. As a result, these two $\nu_{FKM}=1$ systems cannot be distinguished by the values of the line invariants $\gamma_{1}^I \equiv \gamma_2^I \equiv 0$. However, the $\nu_{1x}^I$ are manifestly different. In fact, using the expressions listed in Table~\ref{tab:realspaceinv}  the first Kane-Mele model has $\nu_{1a}^I=\nu_{1b}^I=\nu_{1c}^I \equiv 1$ and $\nu_{1d}^I=0$ while these invariants are reversed by switching the hopping amplitude sign, {\it i.e.} $\nu_{1a}^I=\nu_{1b}^I=\nu_{1c}^I \equiv 0$ and $\nu_{1d}^I=1$. Note that the quantized corner charges in the heterostructure containing both quantum spin-Hall insulators, $Q_{1x}= \Delta \nu^I_{1x} \equiv 1$, are consistent with the direct real-space calculation of Fig.~\ref{fig:qshCC}(c),(d). 

\section{Quantized corner charges as a probe of fragile topology}
Next, we will show that the partial real-space invariants represent a powerful diagnostic tool to detect fragile topological phases. As mentioned above, 
the absence of gapless edges guarantees that 
topological fragile phases do possess quantized corner charges that are in a one-to-one correspondence with the partial real-space invariants. 
In the following, we will derive an inequality that the set of partial real-space invariants necessarily satisfies if the insulator is an atomic one. Consequently, 
a violation of this
inequality indicates that the system 
must 
be a fragile topological insulator as long as its edges are insulating. 

\begin{figure}
\includegraphics[width=1\columnwidth]{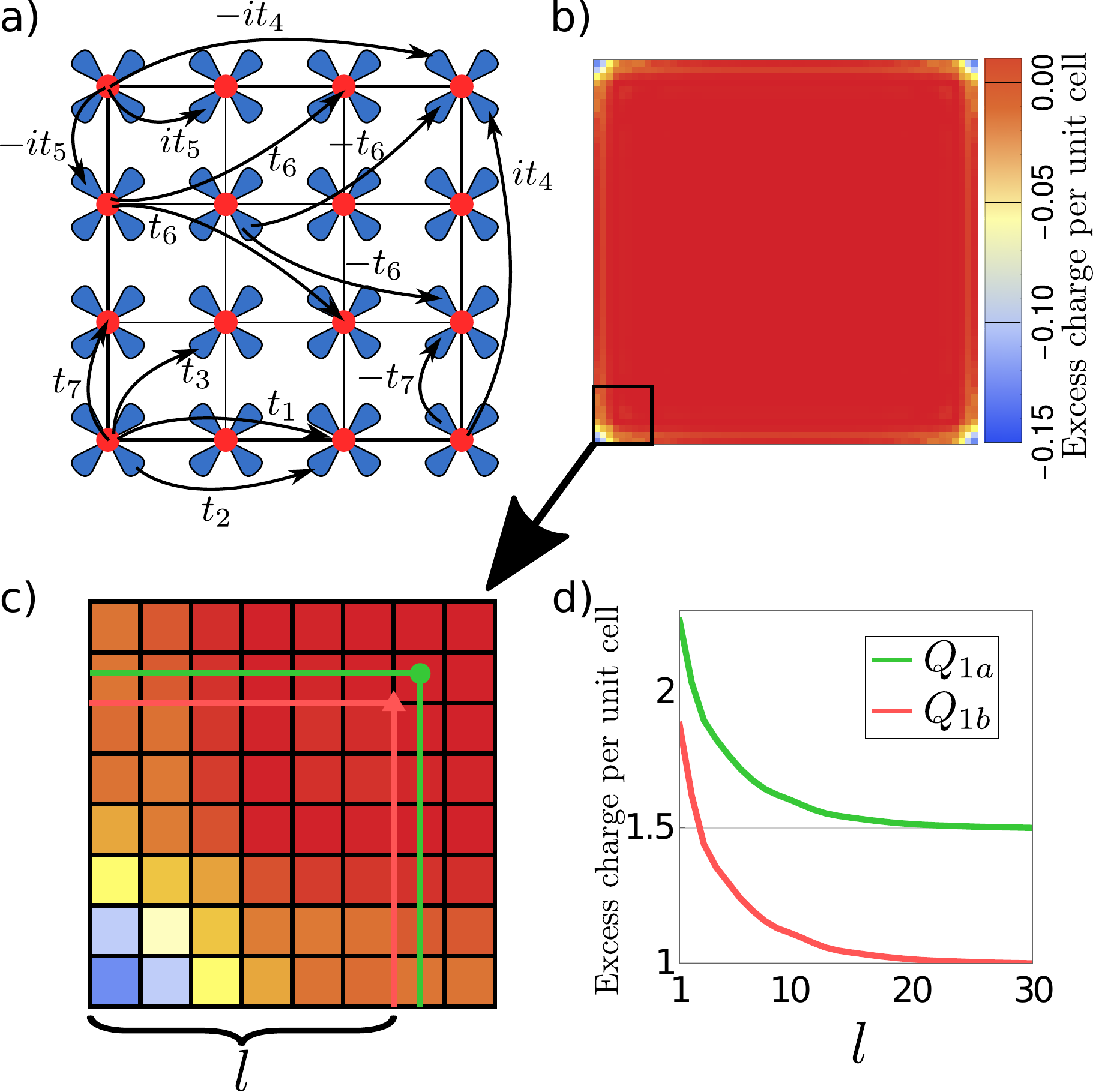}
\caption{(a) Schematic drawing of the hopping parameters for one spin sector of the $\mathcal{C}_4$ symmetric fragile topological insulator model. (b) Charge density of a $\mathcal{C}_4$ symmetric fragile topological insulator for a square sample of 60 by 60 unit cells. (c) Charge density of the lower left corner. The sum of the charge in the red and green areas give the corner charges $Q_{1a}$ and $Q_{1b}$ respectively. (d) Corner charges as a function of the size of the summed region $l$. The corner charges exponentially go to the values of $1.5$ and $1$. Charge density is calculated for $t_1=0.25$ $t_2=-0.4$ $t_3=0.55$ $t_4=-0.35$ $t_5=-0.56$ $t_6=0.125$ $t_7=0.25$. \label{fig:2}}
\end{figure}

Let us first introduce for each special Wyckoff position the positive integer quantities directly related to the partial real-space invariants $\bar{\nu}^I_{m x} = \nu^I_{m x}~\textrm{mod}~(n/m)$. The  $\bar{\nu}^I_{m x}$'s can be immediately related via an exact equality to the quantized part of the corner charges, namely $Q_{m x} = 2 \bar{\nu}^I_{m x} m / n$. Even more importantly, for an insulating phase adiabatically connected to an atomic limit the $\bar{\nu}^{I}_{m x}$ represent lower bounds for the number of exponentially localized Wannier functions with center of charge coinciding with the special Wyckoff positions $N_{m x}  \geq 2~\bar{\nu}^I_{m x}$. Furthermore, the total number of electrons in the unit cell satisfies $N_F  \geq \sum m \times N_{m x}$ where the sum runs  over all the special Wyckoff positions in the unit cell of the rotation symmetric crystal. Combining these two inequalities allows us to derive the following condition fulfilled by a generic atomic insulating phase $N_F \geq \sum 2~m~\times \bar{\nu}^I_{mx}$.  
Insulating phases for which this inequality is violated do not allow an adiabatic deformation to an atomic limit, and hence correspond to fragile topological insulators. 
We can encode this criterion in a discriminant 
$${\mathcal D}_{\mathcal{C}_n} = N_F - \sum 2~m~\times \bar{\nu}^I_{mx},$$
where, as before, the sum runs over all special Wyckoff positions in the unit cell of the rotation-symmetric crystal. 
A negative value of $\mathcal{D}_{\mathcal{C}_n}$ implies 
the existence of a topological obstruction in deforming the insulating phase to an atomic limit, and consequently a fragile topological nature.

We now demonstrate the diagnostic capability of this discriminant, and
consider a concrete 
realization
of a fragile phase in a 
${\mathcal C}_4$-symmetric 
crystal. The model is schematically shown in Fig.~\ref{fig:2}(a) and is 
defined on
a square lattice, with each-unit cell hosting one $s$-like and one $d$-like orbital. 
Besides
conventional nearest-neighbor hoppings we have also included fairly strong long-range hopping processes. At half-filling both the bulk and the edges of this system are completely insulating [see Appendix E], which implies that the corner charges are well-defined
and quantized. 
Explicit computation of the corner charges in the open geometry of Figs.~\ref{fig:2}(b)-(d) yields the following results: 
$Q_{1a}=3/2,\,Q_{1b}=1,\textrm{ and }Q_{2c}=0$. 
This is in agreement with the computation of the bulk partial real-space invariants that can be easily performed decomposing the time-reversal invariant insulator in two time-reversed Chern insulator with channel Chern number $C^{I,II}=\pm 4$ [see Appendix E]. Using the expression for the real-space invariants listed in Table~\ref{tab:realspaceinv}, we indeed find $\bar{\nu}^I_{1a}=3$, $\bar{\nu}^I_{1b}=2$ and $\bar{\nu}^I_{2c}=0$. 
Moreover, since at half-filling only one pair of Kramers related bands are occupied, {\it i.e.} $N_F=2$, we have that the discriminant ${\mathcal D}_{\mathcal{C}_4} = - 8$, thus signaling a fragile topological insulating nature.

We corroborate this finding by verifying the hallmark of fragile topological insulators -- the decay into an atomic insulating state by addition of certain topologically trivial electronic bands. Consider, for instance,  the addition of a single Kramers' related pair of bands. To preserve the fourfold rotation symmetry, the added atomic bands will either correspond to a localized Wannier Kramers' pair centered at the origin of the unit cell $1a$, or at the corner of the unit cell $1b$. In the latter case, we find that $\bar{\nu}^I_{1b}=3$ whereas $\bar{\nu}^I_{1a,2c}$ remain unaffected. This increase in the corner charge $Q_{1b}$ is exactly compensated by the change of  $N_F = 2 \rightarrow 4$ in the discriminant ${\mathcal D}_{{\mathcal C}_4}$ that consequently remains negative indicating that the system is not trivialized.
Let us now consider the addition of a Wannier Kramers pair localized at $1a$. This implies that $\bar{\nu}^I_{1a}$ is modified according to $\bar{\nu}^I_{1a} = 3 \rightarrow 0$ while the other topological crystalline invariants remain unchanged. This consequently implies a change in the discriminant ${\mathcal D}_{\mathcal{C}_4}=-8\rightarrow 0$, which verifies the decay of the fragile topological insulator into a conventional atomic insulator.

It is important to point out that in a ${\mathcal C}_4$ (${\mathcal C}_6$) symmetric crystal the presence of the twofold rotation symmetry allows one to simultaneously define also a ${\mathcal D}_{\mathcal{C}_2}$ discriminant [see Appendix F]. A negative value of  this discriminant 
automatically implies a negative value for ${\mathcal D}_{\mathcal{C}_4}$ (${\mathcal D}_{\mathcal{C}_6}$). 
However, the converse 
needs not to be true. This is verified, for instance, in the model of Fig.~\ref{fig:2}(a) where an electron pair is added at the $1b$ corner of the unit cell. 
The fourfold rotation discriminant ${\mathcal D}_{\mathcal{C}_4}=-8$ implies fragile topology whereas the ${\mathcal C}_2$ specific discriminant ${\mathcal D}_{\mathcal{C}_2}$ identically vanishes and on the contrary would signal an atomic insulator. This property implies that the fragile topology of this model cannot be seen using diagnostic tools specifically designed for twofold rotation symmetric systems, as for instance, the Wilson-loop based indices developed in Refs.~\cite{koo19,wie18,ale14}. Furthermore, the fact that the fragile topology relies purely on the fourfold rotation symmetry, implies that a structural orthorhombic distortion also yields an electronic topological phase transition to an atomic insulating phase. 

Finally, we point out that in ${\mathcal C}_2$-symmetric crystals all insulating phases are atomic for $N_F \geq 6$: this is because $\bar{\nu}^I_{1x}=0,1$ and consequently the discriminant has an upper bound for an even number of Kramers' related pairs of bands ${\mathcal D}_{{\mathcal C}_2} \geq N_F - 8$ while for an odd number of pairs of bands ${\mathcal D}_{{\mathcal C}_2} \geq N_F - 6$. 
On the contrary, the existence of ${\mathcal C}_4$-protected fragile topological phases is not limited to systems with two occupied pairs of Kramers' related bands. 
If starting out from the model in Fig.~\ref{fig:2}(a) we would consider the addition of two Wannier Kramers' pairs centered at the two Wyckoff positions $2c$, the change in $N_F = 2 \rightarrow 6$ would be exactly compensated  by the increase in $\bar{\nu}^I_{2c} = 0 \rightarrow 1$ that leaves the discriminant unaltered. 

\section{Conclusions}
In short, we have introduced the  gauge-invariant crystalline topological indices that govern the quantized corner charges present in two-dimensional rotation symmetric insulators with time-reversal symmetry. 
We dubbed these topological integers partial real-space invariants. They cannot be expressed in terms of  Wilson loop invariants, partial Berry phases or symmetry-based indicators: their computation requires a completely new approach that we have developed throughout our work. Beside defining the bulk-corner correspondence of conventional band insulators adiabatically connected to an atomic limit, the partial real-space invariants can be used to unveil the crystalline topology of quantum spin-Hall insulators, and represent a unique tool to diagnose the recently discovered topological phases of the fragile type in time-reversal symmetric crystals. The bulk-corner correspondence formulated in this work is capable of detecting all fragile topological phases in systems with spin-orbit coupling and  time-reversal symmetry.

\section{Methods} 
\subsection{Constructing the $\mathcal{C}_2$ \& $\Theta$-symmetric gauge along the boundary of the effective Brillouin zone}
\label{sec:constructC2thetaboundary}

In order to construct a continuous, $\mathcal{C}_2$- and $\Theta$-symmetric gauge along the boundary of the effective Brillouin zone, we 
develop a procedure inspired by the
parallel-transport 
procedure 
developed in Ref.~\cite{sol12}. The procedure can be divided in  
three
steps and is 
sketched
in Fig.~\ref{fig:1Dproc}. 
 \\ \\
 \textit{Step 1:} 
  At
 the four high-symmetry momenta $HS\in\{\Gamma=(0,0), Y=(0,\pi),M=(\pi,\pi)X = (\pi,0)\}$ we 
 pick $N_F$ Bloch states
  $|\chi_{t.s.}(HS)\rangle$ 
 in a gauge
 that is $\mathcal{C}_2$ and $\Theta$-symmetric. In the following, we refer to these states at 
 the high-symmetry 
 points of the EBZ
  as target states. 
 We point out that such a set of states can be easily constructed
 by diagonalizing the $\mathcal{C}_2$ symmetry operator.
 \\ \\
\textit{Step 2:} Having selected
the target states at the four high-symmetry 
points in the Brillouin zone,
we next need to find
a continuous gauge that
joins the target states along the upper half of $\partial EBZ$ 
while simultaneously preserving the off-diagonal structure of the ${\mathcal C}_2 \Theta$ sewing matrix.
Let us first consider the line connecting $\Gamma - Y$, and define
an equally spaced mesh $\vec{q}_j=(0,q_{j, y})$ 
with
$j=0,\ldots,\mu$
such that
$q_{0,y} = 0$ and $q_{\mu,y} = \pi$. 
We initialize the parallel transported states by defining the $N_F$ states at $\Gamma$ as:
\begin{align*}
|\chi_{p.t.}(\vec{q}_0)\rangle:= |\chi_{t.s.}(\Gamma)\rangle.
\end{align*}
We can then define the parallel transported states over the entire mesh using the following iterative procedure:
starting from the
parallel-transported 
states
$|\chi_{p.t.}(\vec{q}_j)\rangle$, we 
uniquely determine
the parallel-transported states at $|\chi_{p.t.}(\vec{q}_{j+1})\rangle$ 
by requiring
the overlap matrix ${\mathcal M}(\vec{q}_j,\vec{q}_{j+1}) = \langle\chi_{p.t.}(\vec{q}_j)|\chi_{p.t.}(\vec{q}_{j+1})\rangle$ 
to be
Hermitian and with only positive eigenvalues. 
This can be accomplished by employing a singular value decomposition. Starting from an  
arbitrary gauge 
at $\vec{q}_{j+1}$,  
we write the overlap matrix $\widetilde{{\mathcal M}} (\vec{q}_j,\vec{q}_{j+1}) = \langle\chi_{p.t.}(\vec{q}_j)|\chi(\vec{q}_{j+1})\rangle$ as $\widetilde{{\mathcal M}} = V\Sigma W^\dagger$,  
with $V$ and $W$ unitary matrices and $\Sigma$ a postive real diagonal matrix. 
The parallel-transported states are defined by the unitary transformation of the $ |\chi(\vec{q}_{j+1})\rangle$ states:
\begin{align*}
|\chi_{p.t.}(\vec{q}_{j+1})\rangle = WV^\dagger |\chi(\vec{q}_{j+1})\rangle.
\end{align*}
This guarantees that
the new overlap matrix ${\mathcal M} = V\Sigma V^\dagger$
is Hermitian and positive. This iterative procedure can be repeated
until we have arrived at $(0,\pi) = Y$. 
Moreover, the parallel transport procedure ensures that the block diagonal form of the ${\mathcal C}_2 \Theta$ sewing matrix, or equivalently the constraint  $\ket{\chi^{II}_{p.t.;m}(\vec{q}_{j})} = {\mathcal C}_2 \Theta \ket{\chi^{I}_{p.t.;m}(\vec{q}_{j})}$, is satisfied along the entire line [this is shown in Appendix G].

This, however, is not yet the end of the story: 
by definition,
the parallel transported states at $\Gamma$ coincide with the target state. The same, however,  does not hold true at the $Y$ point:
 the parallel-transported states  $|\chi_{p.t.}(Y)\rangle$ 
will generically not be equal to the target states selected at $Y$. 
In order to correct this, we can in principle apply a residual gauge transformation $U(\vec{q}_j)$
to rotate the parallel-transported states $|\chi_{p.t.}(\vec{q}_j\rangle\rightarrow U(\vec{q}_j)|\chi_{p.t.}(\vec{q}_j)\rangle$.
Specifically, the residual
 gauge transformation should interpolate between the identity matrix at $\Gamma$ and ${\mathcal M}^*_Y = \langle\chi_{t.s}(Y)|\chi_{p.t.}(Y)\rangle^*$ at $Y$. 
 With a continuous and smooth residual gauge transformation, this is only possible if the identity matrix and  ${\mathcal M}^*_Y$
 belong to the same connected component of the subgroup $U(N_F)_{\mathcal{C}_2\Theta}$. 
 Put differently, the determinant of ${\mathcal M}^*_Y$ has to be $+1$. 
 Assuming
  the determinant is instead equal to $-1$, 
 we cannot use any residual gauge transformation to connect our target states at $\Gamma$ and $Y$. However, in this case we have the freedom to
   redefine the target states selected at $Y$, by exchanging a single state from sector $I$ with its Kramers partner in sector $II$. 
  This will switch the sign of the ${\mathcal M}^*_Y$ determinant and eventually allow to interpolate our new target states.
   In concrete terms, we will  apply the following residual gauge transformation:
\begin{align*}
|\chi(\vec{q}_j)\rangle\rightarrow \exp(q_{j,y}\log(M^*_Y)/\pi)|\chi(\vec{q}_j)\rangle.
\end{align*}
To ensure that the states transformed with this additional residual gauge transformation still obey the symmetry constraint $\ket{\chi^{II}_{p.t.;m}(\vec{q}_{j})} = {\mathcal C}_2 \Theta \ket{\chi^{I}_{p.t.;m}(\vec{q}_{j})}$,
we take the logarithm that takes values within the Lie algebra of  $U(N_F)_{\mathcal{C}_2\Theta}$. 
Specifically, denoting
the eigenvalues and eigenvectors of ${\mathcal M}^*_Y$ by $e^{i\theta_j}$ and $v_j$, with $j=1,\ldots,N_F$, one can express this logarithm as:
\begin{align*}
\log({\mathcal M}^*_Y) = \sum_j i\theta_j v_j v^\dagger_j,
\end{align*}
where we require that $\theta_j\in (-\pi,\pi)$, and for simplicity we have assumed that all of the eigenvalues are distinct. 
With this, we finally obtain a continuous gauge along the line connecting $\Gamma$ and $Y$, which gives as output the ${\mathcal C}_2$-symmetric time-reversed target states selected at $\Gamma$ and $Y$ while simultaneously keeping the block off-diagonal structure of the ${\mathcal C}_2 \Theta$ symmetric sewing matrix. 
Next, we repeat the above steps along the line connecting $Y$ and $M$, 
and
the line connecting $M$ and $X$, 
to find our continuous, smooth and symmetric gauge along the upper half of $\partial EBZ.$
\\
\\
 \textit{Step 3:} 
 Finally, we need to
  extend the gauge found at \textit{Step
2} to the entire boundary of the effective Brillouin zone.
Let us first consider the line
connecting $\Gamma$ and $-Y = (0,-\pi)$. 
We can
define $|\chi(\vec{q})\rangle$
for $-\pi<q_{j,y}<0$ as follows: 
\begin{align*}
|\chi_m^\alpha(\vec{q}_j)\rangle & :=-\exp(q_{j,y}(Y_{m}^{\alpha}-\Gamma_{m}^{\alpha})/2)\Gamma_m^\alpha\mathcal{C}_2|\chi_m^\alpha(-\vec{q}_j)\rangle,
\end{align*}
where $Y_m^\alpha$ and $\Gamma_m^\alpha$ denote the $\mathcal{C}_2$ eigenvalues of the target states at $Y$ and $\Gamma$, respectively. Note that the prefactor $-\exp(q_{j,y}(Y_{m}^{\alpha}-\Gamma_{m}^{\alpha})/2)\Gamma_m^\alpha$ ensures that the gauge in the negative half 
of $\partial EBZ$
matches the gauge in the positive half. 
Next, we 
can implement
this procedure 
in an analogous way
along the line connecting $X$ and $-M = (\pi,-\pi)$. The gauge along the 
line
connecting $Y$ and $M$ in the lower half 
can be instead
simply taken to be equal to the section connecting $Y$ and $M$ in the upper half.

\subsection{Computing the Berry phase and ${\mathcal C}_2$ symmetry label contributions to the topological invariants}

With our continuous and ${\mathcal C}_2$ \& $\Theta$--symmetric gauge along the effective Brillouin zone boundary in our hands, we can straightforwardly compute the (partial) Berry phase contribution as well as the multiplicities entering the $\nu^I_{1x}$ expressions of the topological invariants. To compute in particular the Berry phase contribution let
$\vec{q}_j$ with $j=0,\ldots,\mu$
parametrize the mesh along the effective Brillouin zone boundary. Then, using the gauge found in the previous step, 
we have
\begin{align*}
\oint_{\partial EBZ} \mathrm{d}\vec{q}\cdot Tr(\vec{A}^I(\vec{q})) & = \sum_j\Im(\log(\det({\mathcal M}^I(\vec{q}_j,\vec{q}_{j+1}))),
\end{align*}
with ${\mathcal M}^I(\vec{q}_j,\vec{q}_{j+1})_{m,n} = \langle\chi^I_m(\vec{q}_j)|\chi^I_n(\vec{q}_{j+1})\rangle$ the overlap matrix between states at adjacent momenta. Note that this expression does return the Berry phase for the gauge that we have constructed, i.e. the equality is a true equality (not modulo $2\pi$).
\subsection{Constructing a $\mathcal{C}_2\Theta$-symmetric gauge in the effective Brillouin zone} 

In order to obtain a continuous $\mathcal{C}_2\Theta$-symmetric gauge in the effective Brillouin zone, we employ 
again a parallel transport procedure
We initialize our procedure by selecting
at the lower-left corner of the effective Brillouin zone, 
the $-Y=\left(0 , -\pi \right)$ point,
a set of Bloch states such that the $\mathcal{C}_2\Theta$-constraint 
$\ket{\chi^{II}_{p.t.;m}(\vec{q}_{j})} = {\mathcal C}_2 \Theta \ket{\chi^{I}_{p.t.;m}(\vec{q}_{j})}$
is obeyed, see Fig.~\ref{fig:1Dproc}(i). Next, we 
use the iterative
parallel transport 
procedure
in the $\hat{q}_y$-direction to obtain a $\mathcal{C}_2\Theta$-symmetric gauge along the left edge of the effective Brillouin zone, see Fig.~\ref{fig:1Dproc}(j). 
We thereafter use each point along this line
as a starting point to 
iterate the
parallel transport 
of these
Bloch states into the $\hat{q}_x$-direction. This is illustrated in Fig.~\ref{fig:1Dproc}(k) and (d). 
Upon 
completion of these steps, we find
 a continuous $\mathcal{C}_2\Theta$-symmetric 
 gauge in the entire effective Brillouin zone.

\subsection{Computing the ${\mathbb Z}_2$ winding number of ${\mathcal U}_{{\mathcal C}_2 \Theta}(N_F)$}
\label{sec:computeWinding}

To show how to compute the winding number of the overlap matrix $\mathcal U = \braket{\chi(\vec{q})  | \Psi(\vec{q})}$, we 
recall that as shown in Appendix B the group of $\mathcal{C}_2\Theta$-preserving gauge transformations is conjugate to the orthogonal group. 
It is also instructive to remember that the
logarithm of a non-zero complex number $z = \rho e^{i\theta}$, with $\rho\in \mathbb{R}_+$ and $\theta\in\mathbb{R}$ is a multi-valued function:
the logarithm $\log(z)$ is only well-defined up to integer multiples of $ 2\pi i $. 
A simple way to 
define a proper
 single-valued function is to require 
  the imaginary part of the logarithm 
  to take
 values in the open
 interval $(-\pi,\pi]$. One refers to such a logarithm as the principal logarithm. 

We can now analogously define
the
principal
logarithm of an element of the special orthogonal group. First, we consider 
for simplicity
the special orthogonal group in two dimensions. 
Let $M\in SO(2)$, 
and
denote its two eigenvectors 
as
$v_1$ and $v_2$, with 
corresponding
eigenvalues $e^{i\theta_1}$ and $e^{i\theta_2}$. 
Since 
$M$ is a real unitary matrix, 
the following relation must hold
$e^{i\theta_2} = e^{-\i\theta_1}$. 
Moreover, 
the eigenvectors $v_1$ and $v_2$ are related 
to each other
by complex conjugation
(in the rare event that $\theta= 0,\pi$ this might require a basis transformation). 
It is therefore natural to define
the principal logarithm 
of the
matrix $M$ as follows:
\begin{align*}
\log(M) &:= i\theta_1v_1 v_1^\dagger + c.c. = \theta_1 X_1, \textrm{ with }\theta_1 \in (-\pi,\pi], 
\end{align*}
and where
$c.c.$ denotes the complex conjugate. In this way, we can transform
the multi-valued logarithm into a single-valued function
as long as
$\theta_1\neq \pi$. 
In fact, for $\theta_1=\pi$ there is an intrinsic ambiguity in the definition of the principal logarithm since we could freely replace $v_1$ by $v_2$.
We can however remedy to this ambiguity
by  requiring that in the expression for the principal logarithm $v_1 : =(1,+i)^T$. 
With this, we can
conclude that there is a one-to-one mapping between $SO(2)$ and $U(1)$. In particular, this implies that their fundamental 
homotopy
groups coincide: $\pi_1(SO(2)) = \pi_1(U(1)) = \mathbb{Z}$. 
We can straightforwardly determine the
$\mathbb{Z}$-number 
associated to an element of $SO(2)$ on a loop  by counting
the number of times 
$n_+$ the logarithm crosses its branch cut clockwise, and subtracting to it to number of times $n_-$
the branch cut is crossed anti-clockwise.
In practice,
one counts the number of times that the principal logarithm makes a sudden jump by $+2\pi i\sigma_y$ 
and
the number of times 
it jumps
by $-2\pi i\sigma_y$. 
Note that $n_{+}$ and $n_-$ are individually not invariant, as an anti-clock wise crossing can annihilate a clock-wise crossing.

We next generalize the results above
to $SO(N)$ with $N>2$, 
assuming for simplicity $N$ to be even. 
Using the sorted real Schur decomposition discussed in Ref.~\cite{bra02}, 
we
can 
group the eigenvectors and eigenvalues into $N/2$ pairs $v_{2j-1}$ and $v_{2j}$, 
and corresponding
eigenvalues $e^{\pm i\theta_j}$. 
With this, we thereafter
define the  principal logarithm as:
\begin{align*}
\log(M) &:= \sum_{j=1}^{N/2} i\theta_j v_{2j-1}v_{2j-1}^\dagger + c.c. = \sum_{j=1}^{N/2} \theta_j X_j,
\end{align*}
where $X_j$ is a skew-symmetric matrix with $\langle\langle X_i,X_j \rangle\rangle =\Tr\left(X_i^T X_j\right) = 2\delta_{i,j}$, and $\theta_j\in(-\pi,\pi]$ for $j=1,\ldots,N/2$. 
Precisely as for the $SO(2)$ case, the principal logarithm is uniquely defined as long as $\theta_j\neq \pi \,\, \forall j$. However, in this case we cannot resolve 
the ambiguity  if
one of the angles $\theta_j = \pi$. 
This difference can be understood as follows.
 A  $n$-dimensional
 rotation
with $n>2$ can be represented 
as for the two-dimensional case with
 pairs of oriented planes, 
here given
 by $X_j$, and corresponding angles $\theta_j$. In the two-dimensional case, 
it is possible
to resolve the ambiguity for $\theta = \pi$, 
since the orientation of a two-dimensional plane can be globally specified.
This does not hold for higher-dimensional rotation because the oriented planes can be rotated. Such a detail
has major consequences for the fundamental 
homotopy
group of $SO(N)$ for $N>2$. Namely, we can no longer distinguish between a logarithm that crosses the branch cut in a clock-wise or anti clockwise direction. Instead, we can only consider the total number of crossings $n$, which is 
a quantity determined up to multiples of two since, as mentioned above crossings can be annihilated in pairs. Put in simple terms,
the fundamental group of $SO(N)$ is given by $\mathbb{Z}_2$ if $N>2$. 

We now present an explicit numerical recipe to determine whether or not a loop in $SO(N)$ is null-homotopic. 
Assume that we are given a 
set
set of orthogonal matrices $M(i)$ with  $i = 1,\ldots, L$,
satisfying the constraint
$\lVert M(i) - M(i+1)\rVert\ll 1$. Furthermore, we require that the set of matrices form a closed loop, i.e. $M(L)=M(1)$. As a first step, we compute the principal logarithm for each of the $L$ matrices, 
ensuring that $\log(M(L)) = \log(M(0))$ to respect the periodicity.
We can
compute the principal logarithm using the following 
Python
code: 
\\

\begin{spverbatim}
import numpy as np
from scipy.linalg import schur, eigvals, expm

def principalLog(M):
    """
    :input M: special orthogonal matrix of even dimension Nf
    :output X0: skew-symmetric matrix of dimension N x N, s.t. exp(X0)=M, and X0 = theta_i * X_i, with theta_i in [-pi,pi]"""
    T, Z = schur(M)
    sort_real_schur(Z, T, 1.,0,inplace=True)
    Nf = len(M)
    X0 = np.zeros((Nf,Nf))
    for i in range(Nf//2):
        u = Z[::, 2*i:2*i + 1]
        v = Z[::,2*i+1:2*i+2]
        x = u * v.T - v * u.T
        theta = np.arctan2(T[2*i,2*i+1],\
        T[2*i,2*i])
        X0 += theta * x.copy()
    return X0  
 
\end{spverbatim}

Here, we used the function sort\_real\_schur, an implementation of the real Schur decomposition of Ref.~\cite{bra02} which can be found at https://gist.github.com/fabian-paul/14679b43ed27aa25fdb8a2e8f021bad5.

Having computed the principal logarithms, we next need to count the total number of branch cut crossings.
To this end,
  we construct a function that returns the number of crossings between two nearby orthogonal matrices. Specifically, it uses that for two nearby orthogonal matrices $\lVert \log(M(i)) - \log(M(i+1))\rVert^2 \approx n\times 8\pi^2$, with $n$ the number of crossings in between. Typically $n$ will be equal to $0$ or $1$. 
This can be implemented in Python as follows:
\begin{spverbatim}
def crossingIndicator(M1,M2):
    """M1 and M2 two nearby orthogonal matrices. Returns n if n 2x2 blocks of the Schur decomposition cross the branch-cut of the logarithm. 
    """
    L1 = principalLog(M1)
    L2 = principalLog(M2)
    delta = L2-L1
    k = np.linalg.norm(delta/(2 * np.pi),ord='fro')**2/2
    k = round(k)
    return k
    
\end{spverbatim}

Finally, we need to sum the number of crossing over all neighboring points along the mesh. Here, we 
can use
the following function:
\begin{spverbatim}
def Z2(listM):
    """listM is a list of orthogonal matrices M(i). 
    Returns the parity of the number of times the branch-cut is crossed"""
    nu = 0
    for i in range(len(listM)-1):
     nu+= crossingIndicator(listM[i],listM[i+1])
    nu+=crossingIndicator(listM[-1],listM[0])
    return nu
\end{spverbatim}

\section*{Acknowledgements}
\begin{acknowledgments}
We thank E.P. van den Ban for useful discussions. C.O. acknowledges support from a VIDI grant (Project 680-47-543) financed by the Netherlands Organization for Scientific Research (NWO). This work is part of the research programme of the Foundation for Fundamental Research on Matter (FOM), which is part of the Netherlands Organization for Scientific Research (NWO). S.K. acknowledges support from a NWO-Graduate Program grant. 
\end{acknowledgments}

\begin{appendix}

\section*{Appendix A: Derivation of the corner charge bulk invariants}

In this appendix we derive the bulk invariants that determine the
corner charge for $\mathcal{C}_{2},\mathcal{C}_{4}$ and $\mathcal{C}_{6}$, first without time-reversal symmetry. Here, we follow the derivation of Ref.~\cite{mie18}.

\subsection*{$\mathcal{C}_{2}$ symmetry}

With $\mathcal{C}_{2}$ symmetry there are four distinct maximally
symmetric Wyckoff positions. For an atomic insulator with one electron
in the unit cell, this electron has to be localized at one of these
Wyckoff positions, and the corresponding Wannier function will have
eigenvalue $\pm i$. A general atomic insulator is characterized by
the eight numbers $N_{1x,\pm i}$, the number of electrons localized
at Wyckoff position $1x$ with wavefunction eigenvalue $\pm i$. 

However, not all of these integers are independent. In particular,
taking an electron with eigenvalue $+i$ and another with $-i$ at
the same Wyckoff position, one can find linear combinations of the
wavefunctions that can adiabatically be moved away in opposite directions
without breaking the symmetry. Hence, the topological invariants are
$\nu_{1x}=-N_{1x,i}+N_{1x,-i}$, which are immune to the moving of
even and odd pairs combinations.

In Table~\ref{tab:c2} we list the possible atomic orbitals (called
elementary band representations), and the corresponding symmetry eigenvalues
in momentum space. Our goal is to find an expression for $\nu_{1x}$
in terms of the symmetry eigenvalues. To do so, we make the ansatz
that the $\nu_{1x}$ can be expressed as a linear combination of the
symmetry eigenvalues,
\begin{align*}
\nu_{1x} & =\alpha_{1,1x}\Gamma_{i}+\alpha_{2,1x}\Gamma_{-i}+\alpha_{3,1x}X_{-i}\\
 & +\alpha_{4,1x}Y_{-i}+\alpha_{5,1x}M_{-i}.
\end{align*}
Combining this ansatz with Table~\ref{tab:c2}, we find a set of linear equations
that can be solved for $\nu_{1x}$, giving the equations 

\begin{align}
\nu^I_{1a} &=\Gamma^I_{i}+\frac{1}{2}[\Gamma^I_{-i} - X^I_{-i}-Y^I_{-i}-M^I_{-i}], \nonumber \\
\nu^I_{1b} &=-\frac{1}{2}[\Gamma^I_{-i} - X^I_{-i}+Y^I_{-i}-M^I_{-i}],\nonumber \\
\nu^I_{1c} &=-\frac{1}{2}[\Gamma^I_{-i} +X^I_{-i}-Y^I_{-i}-M^I_{-i}],\nonumber \\
\nu^I_{1d} &=-\frac{1}{2}[\Gamma^I_{-i} - X^I_{-i}-Y^I_{-i}+M^I_{-i}], \label{eq:c2noChern}
\end{align}

\subsection*{$\mathcal{C}_{4}$ symmetry}

With $\mathcal{C}_{4}$ symmetry there are three symmetric Wyckoff
positions, two of them have site-symmetry group $\mathcal{C}_{4}$,
$1a$ and $1b$, and one has site-symmetry group $\mathcal{C}_{2}$,
$2c$. For position $2c$ we can immediately define
\begin{align*}
\nu_{2c} & =-N_{2c,i}+N_{2c,-i},
\end{align*}
but for the $\mathcal{C}_{4}$-symmetric Wyckoff positions there are
more possible eigenvalues. For these positions ($1a$ and $1b$),
we solve for 
\begin{align*}
\nu_{1x} & =-3N_{1x,e^{i\pi/4}}+N_{1x,e^{-i\pi/4}}+N_{1x,e^{i3\pi/4}}+N_{1x,e^{-i3\pi/4}},
\end{align*}
for $x=a,b.$ Note that this formulation is again immune to moving
away a quartet of electrons at the same position, each with distinct
eigenvalue, which is adiabatically possible in $\mathcal{C}_{4}$
crystals.

In the definition of $\nu_{1x}$, it is arbitrary which term has coefficient $-3$ instead
of $1$ in front of it, and we could have made a different choice. What matters for
our purposes is that $\nu_{1x}\mod4=N_{1x}\mod4$, where $N_{1x}$
denotes the total number of electrons localized at $1x$. This ensures
that the corner charge $Q_{1x}=N_{1x}/2\mod2$, is completely determined
by $\nu_{1x}$.

Having established the invariants that we want to express in terms
of symmetry eigenvalues, we next make the ansatz that they can be
expressed as a linear combination of symmetry eigenvalues. We then
look at Table \ref{tab:c4}, where the elementary band representations
of $\mathcal{C}_{4}$ are listed. This again provides us with a set
of linear equations that we solve to find 

\begin{align}
\nu_{1a}^{I} & =\left(-3\Gamma_{e^{i\pi/4}}^{I}-\frac{3}{2}\Gamma_{e^{i3\pi/4}}^{I}-\Gamma_{e^{-i3\pi/4}}^{I}\right.\nonumber\\
 & \left.-\frac{3}{2} \Gamma_{e^{-i\pi/4}}^{I}+\frac{3}{2}M_{e^{i3\pi/4}}^{I}+2M_{e^{-i3\pi/4}}^{I}+\frac{3}{2}M_{e^{-i\pi/4}}^{I}+X_{-i}^{I}\right).\nonumber\\
\nu_{1b}^{I} & =\left(\frac{3}{2}\Gamma_{e^{i3\pi/4}}^{I}+2\Gamma_{e^{-i3\pi/4}}^{I}+\frac{3}{2}\Gamma_{e^{-i\pi/4}}^{I}\right.\nonumber\\
 & \left.-\frac{1}{2}M_{e^{i3\pi/4}}^{I}-2M_{e^{-i3\pi/4}}^{I}-\frac{1}{2}M_{e^{-i\pi/4}}^{I}-X_{-i}^{I}\right)\nonumber\\
\nu_{2c}^{I} & =\frac{1}{2}\left(\Gamma_{e^{i3\pi/4}}^{I}+\Gamma_{e^{-i\pi/4}}^{I}-M_{e^{i3\pi/4}}^{I}-M_{e^{-i\pi/4}}^{I}\right),
\label{eq:c4noChern}
\end{align}

\subsection*{$\mathcal{C}_{6}$ symmetry}

In the case of $\mathcal{C}_{6}$ symmetry, there are three symmetric
Wyckoff positions, $1a$, $2b$ and $3c$, which have site-symmetry
group $\mathcal{C}_{6}$, $\mathcal{C}_{3}$ and $\mathcal{C}_{2}$.
The corresponding invariant for $1a$ is 
\begin{align*}
\nu_{1a} & =-5N_{1a,e^{i\pi/6}}+\sum_{l}N_{1a,l},
\end{align*}
with $l=e^{i\pi/2},e^{i5\pi/6},e^{-i5\pi/6},e^{-i\pi/2},e^{-i\pi/6}$,
and for $2b$ and $3c$
\begin{align*}
\nu_{2b}= & -2N_{2b,e^{i\pi/3}}+N_{2b,e^{-i\pi/3}}+N_{2b,-1,}\\
\nu_{3c}= & -N_{3c,i}+N_{3c,-i}.
\end{align*}
Following the same procedure, we make the ansatz that these invariants
can be expressed as a linear combination of the symmetry eigenvalues
in momentum space. Consulting Table \ref{tab:c6} of elementary band
representations we then find a set of linear equations we solve for
the linear coefficients, resulting in 

\begin{align}
\nu_{1a}^{I} & =\left(-5\Gamma_{e^{i\pi/6}}^{I}-\frac{5}{2}\Gamma_{e^{i\pi/2}}^{I}-\Gamma_{e^{i5\pi/6}}^{I}-\frac{1}{2}\Gamma_{e^{-i5\pi/6}}^{I}\right. \nonumber\\
& \left.-\Gamma_{e^{-i\pi/2}}^{I}-\frac{5}{2}\Gamma_{e^{-i\pi/6}}^{I}+\frac{3}{2}M_{-i}^{I}+2K_{-1}^{I}+2K_{e^{-i\pi/3}}^{I}\right)\nonumber\\
\nu_{2b}^{I} & =\left(\Gamma_{e^{i\pi/2}}^{I}+\Gamma_{e^{i5\pi/6}}^{I}+\Gamma_{e^{-i\pi/2}}^{I}+\Gamma_{e^{-i\pi/6}}^{I}-K_{-1}^{I}-K_{e^{-i\pi/3}}^{I}\right)\nonumber\\
\nu_{3c}^{I} & =\frac{1}{2}\left(\Gamma_{e^{i\pi/2}}^{I}+\Gamma_{e^{-i5\pi/6}}^{I}+\Gamma_{e^{-i\pi/6}}^{I}-M_{-i}^{I}\right)
\label{eq:C6BulkInv}
\end{align}

\subsection*{Time-reversal symmetric invariants}

In the main text we have seen that for time-reversal symmetric systems, the partial real-space invariants can be calculated in one 
$\mathcal{C}_n$-symmetric channel to give the corner charge. In this case, however, a contribution from the Chern number calculated within the same channel should be added. In the main text we have seen how this comes about for $\mathcal{C}_2$ symmetry, and here we extend this reasoning to show the contributions of the sector Chern number $C^I$ also appear in the topological invariants of time-reversal symmetric systems with $\mathcal{C}_4$ and $\mathcal{C}_6$ symmetry.

Let us begin with $\mathcal{C}_4$ symmetry, and take a Chern insulator with one occupied band and $\mathcal{C}_4$ eigenvalues $\{e^{i\pi/4},e^{-i\pi/4},-i\}$ at $\{\Gamma,M,X\}$, and Chern number $C=1$. To construct a QSH insulator, we take this Chern insulator, and add its time-reversal partner, which will have Chern number $C=-1$ and eigenvalues $\{e^{-i\pi/4},e^{i\pi/4},i\}$ at $\{\Gamma,M,X\}$.

Taking two identical copies of this QSH insulator will result in a system with a trivial Fu-Kane-Mele invariant, and hence does not feature protected edge states. We can now construct two different symmetric sectors. First, we can take the spin-up channel of the first QSHI and the spin-down of the second, resulting in $C^I=0$ and eigenvalues $\{\{e^{i\pi/4},e^{-i\pi/4}\},\{e^{-i\pi/4},e^{i\pi/4}\},\{-i,i\}\}$. Instead we could construct a gauge where spin-up channel of the first QSHI and the spin-up of the second QSHI are in the same sector. In this case we find $C^I=2$ and $\{\{e^{i\pi/4},e^{i\pi/4}\},\{e^{-i\pi/4},e^{-i\pi/4}\},\{-i,-i\}\}$. In the first decomposition we find $\nu^I_{1a}=-2$, $\nu^I_{1b}=0$ and $\nu^I_{1c}=0$, while the second decomposition, which has sector Chern number $C^I=2$, we find $\nu^I_{1a}=-1$, $\nu^I_{1b}=-3$ and $\nu^I_{1c}=-1$. The topological invariants cannot change under the gauge transformation of picking a different decomposition, and hence we see that the expressions for $\nu^I_{1a}$, $\nu^I_{1b}$ each should be amended with a term $-C^I/2$, and $\nu^I_{2c}$ with $C^I/2$.

Let us next consider $\mathcal{C}_6$ symmetry, and take a Chern insulator with Chern number 1 and symmetry eigenvalues $\{e^{i \pi /2},e^{i\pi/3},i\}$ at $\{\Gamma,K,M\}$. By adding a time-reversal copy of this system, with Chern number $-1$ and eigenvalues $\{e^{-i \pi /2},e^{-i\pi/3},-i\}$, we construct a QSHI. Taking two copies of this QSHI, we again construct a system with trivial Fu-Kane-Mele invariant, and no protected edge states.

Constructing first a sector of the spin-up channel of the first, and spin-down channel of the second QSHI, we find a sector with $C^I=0$ and eigenvalues $\{\{e^{i \pi /2},e^{-i \pi /2}\},\{e^{i\pi/3},e^{-i\pi/3}\},\{i,-i\}\}$. Picking instead the spin-up channels of both copies, we construct a sector with $C^I=2$ and eigenvalues $\{\{e^{i \pi /2},e^{i \pi /2}\},\{e^{i\pi/3},e^{i\pi/3}\},\{i,i\}\}$. The first choice, with $C^I=0$ yields $\nu_{1a}=0$, $\nu_{2b}=1$ and $\nu_{3c}=0$, while the second choice, with $C^I=2$ leads to $\nu_{1a}=-5$, $\nu_{2b}=2$ and $\nu_{3c}=1$. To make these quantities invariant under gauge transformations, we thus need again need to add a term $-C^I/2$ to $\nu^I_{1a}$ and $\nu_{3c}$, and a term $C^I$ to $\nu^I_{2b}$.

\begin{table*}
\begin{tabular}{|c|c||c|c|c|c|c|}
\hline 
$\mathcal{C}_{2}$ & $\lambda$ & $\Gamma_{i}$ & $\Gamma_{-i}$ & $X_{-i}$ & $Y_{-i}$ & $M_{-i}$\tabularnewline
\hline 
\hline 
$1a$ & $i$ & 1 & 0 & 0 & 0 & 0\tabularnewline
\hline 
 & $-i$ & 0 & 1 & 1 & 1 & 1\tabularnewline
\hline 
$1b$ & $i$ & 1 & 0 & 1 & 0 & 1\tabularnewline
\hline 
 & $-i$ & 0 & 1 & 0 & 1 & 0\tabularnewline
\hline 
$1c$ & $i$ & 1 & 0 & 0 & 1 & 1\tabularnewline
\hline 
 & $-i$ & 0 & 1 & 1 & 0 & 0\tabularnewline
\hline 
$1d$ & $i$ & 1 & 0 & 1 & 1 & 0\tabularnewline
\hline 
 & $-i$ & 0 & 1 & 0 & 0 & 1\tabularnewline
\hline 
\end{tabular}

\caption{Table of elementary band representations for crystals with $\mathcal{C}_2$ symmetry. \label{tab:c2}}
\end{table*}

\begin{table*}
\begin{tabular}{|c|c||c|c|c|c|c|c|c|c|}
\hline 
$\mathcal{C}_{4}$ & $\lambda$ & $\Gamma_{e^{i\pi/4}}$ & $\Gamma_{e^{i3\pi/4}}$ & $\Gamma_{e^{-i3\pi/4}}$ & $\Gamma_{e^{-i\pi/4}}$ & $M_{e^{i3\pi/4}}$ & $M_{e^{-i3\pi/4}}$ & $M_{e^{-i\pi/4}}$ & $X_{-i}$\tabularnewline
\hline 
\hline 
$1a$ & $e^{i\pi/4}$ & 1 & 0 & 0 & 0 & 0 & 0 & 0 & 0\tabularnewline
\hline 
 & $e^{i3\pi/4}$ & 0 & 1 & 0 & 0 & 1 & 0 & 0 & 1\tabularnewline
\hline 
 & $e^{-i3\pi/4}$ & 0 & 0 & 1 & 0 & 0 & 1 & 0 & 0\tabularnewline
\hline 
 & $e^{-i\pi/4}$ & 0 & 0 & 0 & 1 & 0 & 0 & 1 & 1\tabularnewline
\hline 
$1b$ & $e^{i\pi/4}$ & 1 & 0 & 0 & 0 & 0 & 1 & 0 & 1\tabularnewline
\hline 
 & $e^{i3\pi/4}$ & 0 & 1 & 0 & 0 & 0 & 0 & 1 & 0\tabularnewline
\hline 
 & $e^{-i3\pi/4}$ & 0 & 0 & 1 & 0 & 0 & 0 & 0 & 1\tabularnewline
\hline 
 & $e^{-i\pi/4}$ & 0 & 0 & 0 & 1 & 1 & 0 & 0 & 0\tabularnewline
\hline 
$2c$ & $i$ & 1 & 0 & 1 & 0 & 1 & 0 & 1 & 1\tabularnewline
\hline 
 & $-i$ & 0 & 1 & 0 & 1 & 0 & 1 & 0 & 1\tabularnewline
\hline 
\end{tabular}

\caption{Table of elementary band representations for crystals with $\mathcal{C}_4/\mathcal{S}_4$ symmetry.\label{tab:c4}}
\end{table*}

\begin{table*}
\begin{tabular}{|c|c||c|c|c|c|c|c|c|c|c|}
\hline 
$\mathcal{C}_{6}$ & $\lambda$ & $\Gamma_{e^{i\pi/6}}$ & $\Gamma_{e^{i\pi/2}}$ & $\Gamma_{e^{i5\pi/6}}$ & $\Gamma_{e^{-i5\pi/6}}$ & $\Gamma_{e^{-i\pi/2}}$ & $\Gamma_{e^{-i\pi/6}}$ & $M_{-i}$ & $K_{-1}$ & $K_{e^{-i\pi/3}}$\tabularnewline
\hline 
\hline 
$1a$ & $e^{i\pi/6}$ & 1 & 0 & 0 & 0 & 0 & 0 & 0 & 0 & 0\tabularnewline
\hline 
 & $e^{i\pi/2}$ & 0 & 1 & 0 & 0 & 0 & 0 & 1 & 1 & 0\tabularnewline
\hline 
 & $e^{i5\pi/6}$ & 0 & 0 & 1 & 0 & 0 & 0 & 0 & 0 & 1\tabularnewline
\hline 
 & $e^{-i5\pi/6}$ & 0 & 0 & 0 & 1 & 0 & 0 & 1 & 0 & 0\tabularnewline
\hline 
 & $e^{-i\pi/2}$ & 0 & 0 & 0 & 0 & 1 & 0 & 0 & 1 & 0\tabularnewline
\hline 
 & $e^{-i\pi/6}$ & 0 & 0 & 0 & 0 & 0 & 1 & 1 & 0 & 1\tabularnewline
\hline 
$2b$ & $e^{i\pi/3}$ & 1 & 0 & 0 & 1 & 0 & 0 & 1 & 1 & 1\tabularnewline
\hline 
 & $-1$ & 0 & 1 & 0 & 0 & 1 & 0 & 1 & 0 & 1\tabularnewline
\hline 
 & $e^{-i\pi/3}$ & 0 & 0 & 1 & 0 & 0 & 1 & 1 & 1 & 0\tabularnewline
\hline 
$3c$ & $i$ & 1 & 0 & 1 & 0 & 1 & 0 & 2 & 1 & 1\tabularnewline
\hline 
 & $-i$ & 0 & 1 & 0 & 1 & 0 & 1 & 1 & 1 & 1\tabularnewline
\hline 
\end{tabular}

\caption{Table of elementary band representations for crystals with $\mathcal{C}_6$ symmetry.\label{tab:c6}}
\end{table*}

\section*{Appendix B: Proof that the group of $\mathcal{C}_2 \Theta$-preserving gauge transformations is conjugate to the orthogonal group}

In this Appendix we show that the group of $\mathcal{C}_2\Theta$-preserving gauge transformations is conjugate to the orthogonal group  
We start by recalling that
a $\mathcal{C}_2\Theta$-symmetric gauge ensures that $\langle\psi_m^\alpha|\mathcal{C}_2\Theta|\psi_n^\beta\rangle = \sigma_x^{\alpha,\beta}\delta_{m,n}$. 
Under a residual ${\mathcal U}_{{\mathcal C}_2 \Theta}(N_F)$ gauge transformation
the left-hand side of this equation changes into ${\mathcal U}_{{\mathcal C}_2 \Theta}^*\left(\sigma_x\otimes\delta\right) U^\dagger$. This means that the group of residual gauge transformations satisfies the following constraint:
\begin{align*}
\left(\sigma_x\otimes\delta\right){\mathcal U}_{{\mathcal C}_2 \Theta} ^*(\sigma_x\otimes\delta) =  {\mathcal U}_{{\mathcal C}_2 \Theta} . 
\end{align*}
Next, we multiple the left- and right-hand sides from the left with $K$ and from the right with $K^\dagger$, where $K$ is a unitary matrix given by:
\begin{align*}
K&=\frac{1}{\sqrt{2}}\begin{pmatrix}
1&1\\
i & -i
\end{pmatrix}\otimes\delta.
\end{align*}
We find:
\begin{align*}
K {\mathcal U}_{{\mathcal C}_2 \Theta} K^\dagger &= K \left(\sigma_x\otimes\delta\right){\mathcal U}_{{\mathcal C}_2 \Theta}^*(\sigma_x\otimes\delta) K^\dagger = K^* {\mathcal U}_{{\mathcal C}_2 \Theta}^* K^T \\
&= (K {\mathcal U}_{{\mathcal C}_2 \Theta} K^\dagger)^*.
\end{align*}
Hence, $K {\mathcal U}_{{\mathcal C}_2 \Theta} K^\dagger$ is a real unitary matrix, i.e. $K {\mathcal U}_{{\mathcal C}_2 \Theta} K^\dagger\in O(N_F)$. Given a loop that resides in the component with determinant -1, we can check whether it is contractible by a point by considering a loop in $SO(N_F)$, obtained by applying a $\mathcal{C}_2\Theta$-respecting transformation with determinant -1. An example would be the exchange of a pair of bands from sector $I$ to $II$.

\section*{Appendix C: computational procedure for $\mathcal{C}_4$ and $\mathcal{C}_6$ symmetry}

\begin{figure}
\includegraphics[width=1\columnwidth]{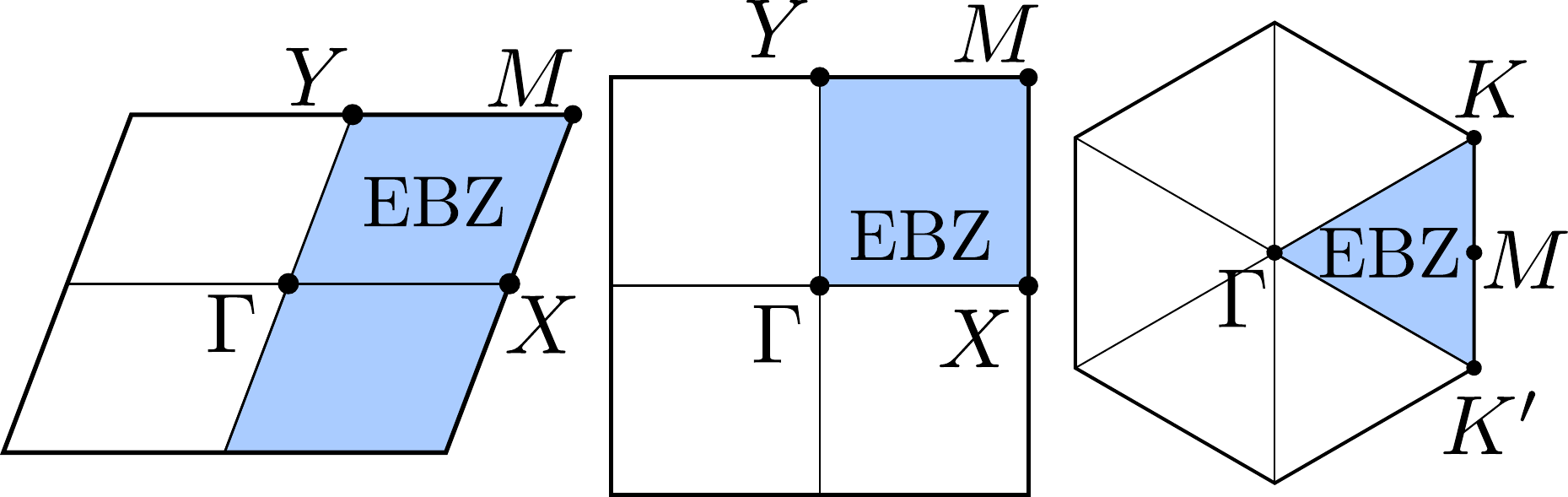}

\caption{Brillouin zones of $\mathcal{C}_{2}$-, $\mathcal{C}_{4}$- and $\mathcal{C}_{6}$-symmetric
crystals. The effective Brillouin zone in each case is the blue
region.\label{fig:EBZ}}

\end{figure}

The procedure for calculating $\nu_{1a}^{I}$, $\nu_{1b}^{I}$ and
$\nu_{2c}^{I}$ for $\mathcal{C}_{4}$-symmetric systems is similar
to that for $\mathcal{C}_{2}$ systems. Since by $\mathcal{C}_{4}$
symmetry, each quadrant of the Brillouin zone is related to each other,
our effective Brillouin zone will now be a quarter of the full Brillouin
zone, $EBZ=[0,\pi]\times[0,\pi]$ (see Fig.~\ref{fig:EBZ})

A set of continuous, periodic and rotation-symmetric projectors $\rho^{I,II}\left(q\right)$
is now equivalent to a set of Bloch waves $|\Psi_{m}^{I,II}\left(\vec{q}\right)\rangle$
in the EBZ such that 
\begin{enumerate}
\item the sewing matrix $S_{\mathcal{C}_{4}}\left(\vec{q}\right)=\langle\Psi_{m}^{\alpha}\left(\mathcal{C}_{4}\vec{q}\right)|\mathcal{C}_{4}|\Psi_{n}^{\beta}\left(\vec{q}\right)\rangle$
is block-diagonal along $\partial EBZ$
\item the sewing matrix $S_{\mathcal{C}_{2}\Theta}\left(\vec{q}\right)=\langle\Psi_{m}^{\alpha}\left(\vec{q}\right)|\mathcal{C}_{2}\Theta|\Psi_{n}^{\beta}\left(\vec{q}\right)\rangle$
is block off-diagonal in the entire EBZ, which in particular it can
be chosen such that $S_{\mathcal{C}_{2}\Theta}\left(\vec{q}\right)_{m,n}^{\alpha,\beta}=\sigma_{x}^{\alpha,\beta}\delta_{m,n}$.
\end{enumerate}
Having such a gauge in the EBZ, the Chern number can be rewritten
using Stokes' theorem, and fourfold rotation symmetry as
\begin{align*}
\frac{C^{I}}{4} & =\frac{1}{2\pi}\ointclockwise_{\partial EBZ}d\vec{q}\cdot Tr\left(\vec{A}^{I}\left(\vec{q}\right)\right).
\end{align*}
Following the same reasoning as for the $\mathcal{C}_{2}$-symmetric
case, we also add the winding number $W\left(U|_{\partial EBZ}\right)$
to make the topological invariants $\nu^{I}$ invariant under arbitrary
$\mathcal{C}_{2}\Theta$-preserving gauge transformations. Hence we
make the substitution 
\begin{align*}
-\frac{C^{I}}{2} & \rightarrow-\frac{1}{\pi}\ointclockwise_{\partial EBZ}d\vec{q}\cdot Tr\left(\vec{A}^{I}\left(\vec{q}\right)\right)-4W\left(U|_{\partial EBZ}\right)
\end{align*}
 in the formulations of the $\nu^{I}$'s listed in Table~\ref{tab:realspaceinv}.

The procedure to compute the topological invariants is then almost
unchanged and consists of the following steps:
\begin{enumerate}
\item construct a continuous $\mathcal{C}_{4}$- and $\theta$-symmetric
gauge of Bloch waves $\Psi\left(\vec{q}\right)$ along $\partial EBZ$,
\item compute the partial Berry phase along $\partial EBZ$, and the multiplicities
of the rotation-symmetry eigenvalues using $\Psi\left(\vec{q}\right)$, 
\item construct a continuous $\mathcal{C}_{2}\theta$-symmetric gauge for
Bloch waves $\chi\left(\vec{q}\right)$ within the EBZ and
\item compute the $\mathbb{Z}_{2}$ or $\mathbb{Z}$ winding number of the
overlap matrix $U\left(\vec{q}\right)$ along $\partial EBZ$. 
\end{enumerate}
To construct a continuous $\mathcal{C}_{4}$- and $\theta$-symmetric
gauge along $\partial EBZ$, we start by picking target states at
the high-symmetry momenta $HS\in\left\{ \Gamma=(0,0),Y=(0,\pi),M=(\pi,\pi),X=(\pi,0)\right\} $,
that are $\theta$-symmetric and $\mathcal{C}_{4}$-symmetric at $\Gamma$
and $M$, and $\mathcal{C}_{2}$-symmetric at $Y$ and $X$. We find
$N_{F}$ of these states $|\chi_{t.s.}\left(HS\right)\rangle$, and
in order to respect $\mathcal{C}_{4}$ symmetry, we define $|\chi_{t.s.}\left(X\right)\rangle=\mathcal{C}_{4}|\chi_{t.s.}\left(Y\right)\rangle$.
Such a gauge can be easily found by diagonalizing the $\mathcal{C}_{4}$
symmetry operator at $\Gamma$ and $M$, and the $\mathcal{C}_{2}$
operator at $Y$.

Having found target states at the high-symmetry points in the BZ,
we next want to find a continuous gauge connecting them, preserving
the off-diagonal structure of the $\mathcal{C}_{2}\theta$ sewing
matrix. To do so we first consider the line $\Gamma-Y$, and follow
the exact same procedure as for $\mathcal{C}_{2}$ symmetry (see the Methods section) to parallel transport the target states from $\Gamma$
to $Y$, and to then apply a residual gauge transformation to ensure
it is smooth. The states along the line $\Gamma-X$ are then defined
as, for $0<q_{j}<\pi$,
\begin{align*}
|\chi_{m}^{\alpha}\left(q_{j},0\right)\rangle & :=\left(\Gamma_{m}^{\alpha}\right)^{-1+q_{j}/\pi}\mathcal{C}_{4}|\chi_{m}^{\alpha}\left(0,q_{j}\right)\rangle.
\end{align*}
Next, using the same procedure we construct a continuous gauge along
the line $X-M$, and then define the states along the line $Y-M$
as, for $0<q_{j}<\pi$, 
\begin{align*}
|\chi_{m}^{\alpha}\left(q_{j},\pi\right)\rangle & :=\left(\Gamma_{m}^{\alpha}\right)^{-1+q_{j}/\pi}\left(X_{m}^{\alpha}\right)^{-q_{j}/\pi}\mathcal{C}_{4}|\chi_{m}^{\alpha}\left(\pi,q_{j}\right)\rangle.
\end{align*}
This procedure gives us a set of states $|\chi\left(\vec{q}\right)\rangle$
along $\partial EBZ$, that are $\mathcal{C}_{4}$ and $\theta$ symmetric,
and that preserve the off-diagonal form of the $\mathcal{C}_{2}\theta$
operator. 

Having found such a symmetric gauge along $\partial EBZ$, we can
calculate the Berry phase contribution, as well as the sector symmetry
eigenvalues in a similar manner. Constructing a $\mathcal{C}_{2}\theta$-symmetric
gauge in the effective Brillouin zone, and consequently computing
the winding number of the overlap matrix $\mathcal{U}=\langle\chi\left(\vec{q}\right)|\Psi\left(\vec{q}\right)\rangle$
is done in the same manner, the only difference being that $\vec{q}$
runs along a different EBZ: a quarter of the full Brillouin zone instead
of half.

Having treated the $\mathcal{C}_{2}$- and $\mathcal{C}_{4}$- symmetric
cases, the $\mathcal{C}_{6}$-symmetric case proceeds along the same lines.
The Brillouin zone is now a hexagon, and the effective BZ runs from
$\Gamma$ to $K$ to $K'$ along $M$, and back to $\Gamma$ (see
Fig.~\ref{fig:EBZ}). A set of continuous, periodic and rotation-symmetric
projectors $\rho^{I,II}\left(q\right)$ is now equivalent to a set
of Bloch waves $|\Psi_{m}^{I,II}\left(\vec{q}\right)\rangle$ in the
EBZ such that 
\begin{enumerate}
\item the sewing matrix $S_{\mathcal{C}_{6}}\left(\vec{q}\right)=\langle\Psi_{m}^{\alpha}\left(\mathcal{C}_{4}\vec{q}\right)|\mathcal{C}_{6}|\Psi_{n}^{\beta}\left(\vec{q}\right)\rangle$
is block-diagonal along $\partial EBZ$
\item the sewing matrix $S_{\mathcal{C}_{2}\Theta}\left(\vec{q}\right)=\langle\Psi_{m}^{\alpha}\left(\vec{q}\right)|\mathcal{C}_{2}\Theta|\Psi_{n}^{\beta}\left(\vec{q}\right)\rangle$
is block off-diagonal in the entire EBZ, which in particular it can
be chosen such that $S_{\mathcal{C}_{2}\Theta}\left(\vec{q}\right)_{m,n}^{\alpha,\beta}=\sigma_{x}^{\alpha,\beta}\delta_{m,n}$.
\end{enumerate}
Having such a gauge, the Chern number can be calculated considering
the Berry phase along $\partial EBZ$, using
\begin{align*}
\frac{C^{I}}{6} & =\frac{1}{2\pi}\ointclockwise_{\partial EBZ}d\vec{q}\cdot Tr\left(\vec{A}^{I}\left(\vec{q}\right)\right).
\end{align*}
Here, again we add the winding number $W\left(U|_{\partial EBZ}\right)$
to the formulas of the $\nu^{I}$ to make the formulation invariant
under $\mathcal{C}_{2}\Theta$-symmetry preserving gauge transformations.
In the formulas of $\nu^{I}$ we thus need to substitute 
\begin{align*}
-\frac{C^{I}}{2} & \rightarrow-\frac{3}{2\pi}\ointclockwise_{\partial EBZ}d\vec{q}\cdot Tr\left(\vec{A}^{I}\left(\vec{q}\right)\right)-6W\left(U|_{\partial EBZ}\right).
\end{align*}

To compute $\nu_{1a}^{I}$,$\nu_{2b}^{I}$ and $\nu_{3c}^{I}$, we
follow the almost identical four steps as for $\mathcal{C}_{4}$ symmetry:
\begin{enumerate}
\item construct a continuous $\mathcal{C}_{6}$- and $\Theta$-symmetric
gauge of Bloch waves $\Psi\left(\vec{q}\right)$ along $\partial EBZ$,
\item compute the partial Berry phase along $\partial EBZ$, and the multiplicities
of the rotation-symmetry eigenvalues using $\Psi\left(\vec{q}\right)$, 
\item construct a continuous $\mathcal{C}_{2}\Theta$-symmetric gauge for
Bloch waves $\chi\left(\vec{q}\right)$ within the EBZ and
\item compute the $\mathbb{Z}_{2}$ or $\mathbb{Z}$ winding number of the
overlap matrix $U\left(\vec{q}\right)=$ along $\partial EBZ$, 
\end{enumerate}
where the difference is that in step 1 we need to construct a $\mathcal{C}_{6}$-
and $\Theta$-symmetric gauge, and the EBZ is different. We start
by constructing $\mathcal{C}_{6}$- and $\theta$-symmetric target
states at $\Gamma,K$ and $M$ by diagonalizing the rotation-symmetry
operators at these points in the BZ and to respect $\mathcal{C}_{6}$
symmetry, we define $|\chi_{t.s.}\left(K'\right)\rangle=\mathcal{C}_{6}|\chi_{t.s.}\left(K\right)\rangle$. 

To find smooth, symmetric states along the $\Gamma-K$ line, we follow
the parallel transport procedure outlined above. The states along
$\Gamma-K'$ are then defined as (for $q_{j}$ along $\Gamma-K$)
\begin{align*}
|\chi_{m}^{\alpha}\left(\mathcal{C}_{6}q_{j}\right)\rangle & :=\left(\Gamma_{m}^{\alpha}\right)^{-1+q_{j}/\pi}\left(K_{m}^{\alpha}\right)^{-q_{j}/\pi}\mathcal{C}_{6}|\chi_{m}^{\alpha}\left(q_{j}\right)\rangle.
\end{align*}
Along the line $M-K$, we again use the parallel transport procedure
to find symmetric and continuous states. This line is related by $\mathcal{\mathcal{C}}_{2}$
symmetry to the line $M-K'$, and hence we define the states along
this line, for $q_{j}$ on the line $M-K$,
\begin{align*}
|\chi_{m}^{\alpha}\left(\mathcal{C}_{2}q_{j}\right)\rangle & :=\left(M_{m}^{\alpha}\right)^{-1+q_{j}/\pi}\left(K_{m}^{\alpha}\right)^{-q_{j}/\pi}\mathcal{C}_{2}|\chi_{m}^{\alpha}\left(q_{j}\right)\rangle.
\end{align*}
Having constructed this symmetric and continuous gauge along $\partial EBZ$,
steps 2-4 are identical to those for $\mathcal{C}_{2}$ and $\mathcal{C}_{4}$
symmetry, where only the EBZ is different. 

\section*{Appendix D: Relating $\mathcal{C}_4$ and $\mathcal{C}_6$ invariants to crystalline line invariants}

In the main text we have seen how the invariants $\nu^I_{1x}$ are related to the partial polarizations $\gamma^I_{1,2}$. In this brief Appendix, we show this relation for $\mathcal{C}_4$ and $\mathcal{C}_6$ symmetry. The results follow from the fact that the $\mathcal{C}_4$ and $\mathcal{C}_6$ eigenvalues directly determine the $\mathcal{C}_2$ eigenvalues.

For the $\mathcal{C}_{4}$-symmetric case we obtain
\begin{align*}
\left(2\nu_{1b}^{I}+\nu_{2c}^{I}\right)\mod2 & =\left(\nu_{FKM}-\gamma_{1}^{I}\right)\mod2,
\end{align*}
Note that for a $\mathcal{C}_{4}$
symmetric system we have $\gamma_{1}^{I}=\gamma_{2}^{I}$. Summing
all three invariants, we find
\begin{align*}
 & 2\left(\nu_{1a}^{I}+\nu_{1b}^{I}+\nu_{2c}^{I}\right)\mod4\\
= & \left(-3\Gamma_{e^{i\pi/4}}^{I}+\Gamma_{e^{i3\pi/4}}^{I}+\Gamma_{e^{-i3\pi/4}}^{I}+\Gamma_{e^{-i\pi/4}}^{I}\right)\mod4\\
= & \left(N_{F}/2\right)\mod4.
\end{align*}
For $\mathcal{C}_{6}$ we obtain
\begin{align*}
\left(3\nu_{1a}+\nu_{3c}\right)\mod2 & =\left(\nu_{FKM}-\gamma_{1}^{I}\right)\mod2,
\end{align*}
 and 
\begin{align*}
3\left(\nu_{1a}^{I}+\nu_{2b}^{I}+\nu_{3c}^{I}\right)\mod6 & =\left(N_{F}/2\right)\mod6.
\end{align*}

The Fu-Kane-Mele invariant $\nu_{FKM}$
is thus equal to the parity of the sector Chern number, while the
partial polarizations $\gamma_{1,2}^{I}$ are easily calculated from
the sector eigenvalues.

\section*{Appendix E: Details of the $\mathcal{\mathcal{C}}_{4}$-symmetric fragile topological insulator}

Here, we provide additional details of the fragile topological insulator
presented in Fig.~5. The model is constructed by considering the Hamiltonian as shown in
Fig.~5(a), consisting of an $s$- and $d$-orbital on a square lattice,
and is a Chern insulator with $C=-4$. This system is time-reversal
symmetry broken, and we add its time-reversal partner such that the
combined system is time-reversal symmetric. Each time-reversal channel
by itself then has Chern number $C=\mp4$. From the way we have constructed
this model, namely taking two systems that have broken TRS and $\mathcal{C}_{4}$
symmetry, we immediately obtain a symmetric gauge, with sector Chern
number $C^{I}=-4$ and sector eigenvalues $\{e^{-i3\pi/4},e^{-i3\pi/4},i\}$
at the high-symmetry points $\{\Gamma,M,X\}$. We note that this decomposition
can only be made if the two time-reversal sectors are not coupled,
that is, there is no spin-orbit coupling. 

To make sure the counter-propagating edge states gap out, we add a
term $-\sin\left(k_{x}\right)\left[\sigma_{y}\otimes\tau_{0}\right]$,
where $\tau_{i}$ and $\sigma_{i}$ are Pauli matrices that act in
orbital and spin space respectively and we take $\lambda/t=-0.35$.
Adding this terms means we do not immediately have a symmetric gauge,
and we have to resort to the methods of Sec.~VIII to calculate the
topological invariants.

\begin{figure}
\includegraphics[width=1\columnwidth]{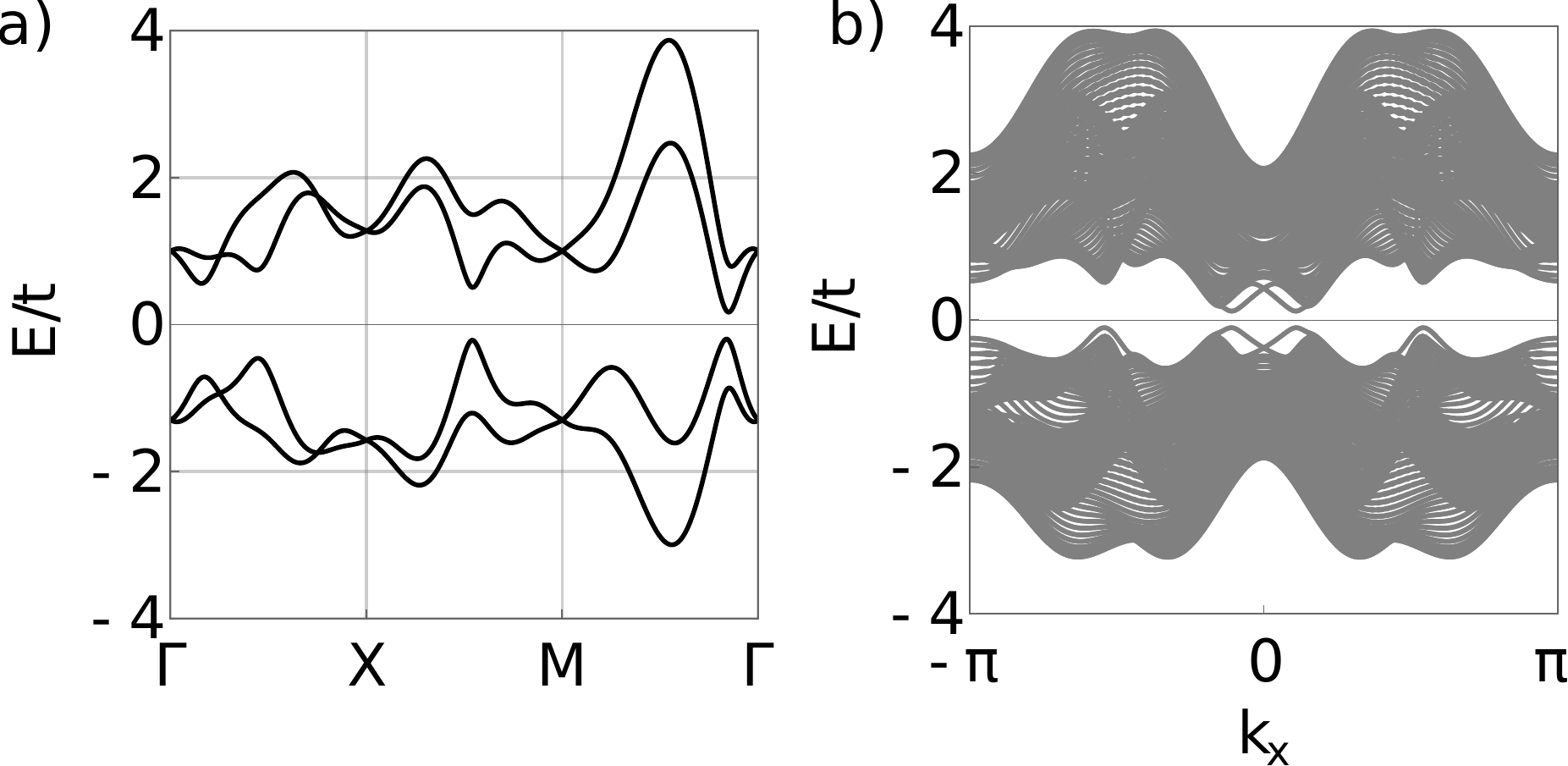}

\caption{(a) Bulk and (b) ribbon spectrum of the $\mathcal{C}_{4}$-symmetric
topological insulator. See Fig.~5 of the main text for the parameter
values.\label{fig:c4fragapp}}
\end{figure}

The bulk and ribbon spectrum (finite along the
$y$-direction) are plotted in Fig.~\ref{fig:c4fragapp}. We see
that they are both gapped, and hence picking a Fermi energy inside
both gaps, the corner charges are well-defined.

\section*{Appendix F: Diagnosis of fragile topology protected by $\mathcal{C}_{2}$ symmetry in
$\mathcal{C}_{4,6}$ symmetric crystals}

In this appendix we show how to compute the ${\mathcal D}_{{\mathcal C}_2}$ discriminant in ${\mathcal C}_{4,6}$ symmetric crystals. This discriminant signals if a fragile topological phase is protected by the twofold rotation symmetry. For ${\mathcal C}_4$ symmetric crystals we have that the corner charge with respect to $1a$
in a $\mathcal{C}_{4}$ symmetric system con be computed by integrating the charge density in a single corner. In order to determine $Q_{1a}$ in a ${\mathcal C}_2$ symmetric corner, 
we need to integrate the charge density over two corners, which are identical by $\mathcal{C}_4$ symmetry. In terms of the ${\bar \nu}_{mx}$ we therefore have 
$$\bar{\nu}_{1a}^{{\mathcal C}_2}  \equiv  \bar{\nu}_{1a}^{{\mathcal C}_4}  \textrm{mod}~2$$
Similarly for the corner charge measured with respect to the corner of the unit cell we have 
$$\bar{\nu}_{1b}^{{\mathcal C}_2}  \equiv  \bar{\nu}_{1b}^{{\mathcal C}_4}  \textrm{mod}~2.$$
On the contrary, for the ${\mathcal C}_2$-symmetric $2c$ Wyckoff positions we have 
 $$\bar{\nu}_{2c}^{{\mathcal C}_2}  \equiv  \bar{\nu}_{2c}^{{\mathcal C}_4} $$

From this it follows that for a $\mathcal{C}_{4}$-symmetric crystal
one can determine the $\mathcal{C}_{2}$ discriminant by calculating
\begin{align*}
\mathcal{D}_{\mathcal{C}_{2}} & =N_{F}-2\times\left[\bar{\nu}_{1a}^{{\mathcal C}_2} + \bar{\nu}_{1b}^{{\mathcal C}_2} + 2\times \bar{\nu}_{2c}^{{\mathcal C}_2}  \right].
\end{align*}
Going back to the $\mathcal{C}_{4}$-symmetric fragile insulator of
Fig.~5, which had $Q_{1a}^{\mathcal{C}_{4}}=3/2$, $Q_{1a}^{\mathcal{C}_{4}}=1$,
$Q_{2c}^{\mathcal{C}_{4}}=0$ and $\mathcal{D}_{\mathcal{C}_{4}}=-8$, we
find $\mathcal{D}_{\mathcal{C}_{2}}=0.$ Hence the $\mathcal{C}_{2}$
discriminant does not diagnose the fragile topological phase, while
the $\mathcal{C}_{4}$ discriminant does.
Using the same reasoning for $\mathcal{C}_{6}$-symmetric systems, we have 
$$\bar{\nu}_{1a}^{{\mathcal C}_2}  \equiv  \bar{\nu}_{1a}^{{\mathcal C}_6}  \textrm{mod}~2,$$
and 
$$\bar{\nu}_{3c}^{{\mathcal C}_2}  \equiv  \bar{\nu}_{3c}^{{\mathcal C}_6}.$$
Consequently, the discriminant signalling fragile topological phases protected by the twofold rotation symmetry reads 
\begin{align*}
\mathcal{D}_{\mathcal{C}_{2}} & =N_{F}-2\times\left[\bar{\nu}_{1a}^{{\mathcal C}_2} + 3\times\bar{\nu}_{3c}^{{\mathcal C}_2}  \right].
\end{align*}

\subsection*{Appendix G: Parallel transport and the ${\mathcal C}_2 \Theta$ sewing matrix}

Here, we show that the parallel transport procedure ensures that the block diagonal form of the ${\mathcal C}_2 \Theta$ sewing matrix, or equivalently the constraint  $\ket{\chi^{II}_{p.t.;m}(\vec{q}_{j})} = {\mathcal C}_2 \Theta \ket{\chi^{I}_{p.t.;m}(\vec{q}_{j})}$, is satisfied along the entire line.
We consider two neighboring momenta $\vec{q}_j$ and $\vec{q}_{j+1}$. We find the following relation:
\begin{align*}
{\mathcal M} (\vec{q}_j,\vec{q}_{j+1}) &= \langle \chi^\alpha_{p.t.;m}(\vec{q}_{j})|\chi^\beta_{p.t.;n}(\vec{q}_{j+1})\rangle\\
&=\langle\mathcal{C}_2\Theta \chi^\beta_{p.t.;n}(\vec{q}_{j+1})|\mathcal{C}_2\Theta\chi^\alpha_{p.t.;m}(\vec{q}_j)\rangle\\
&=\mathcal{S}^\dagger_{\mathcal{C}_2\Theta}(\vec{q}_{j+1}){\mathcal M}^\dagger(\vec{q}_j,\vec{q}_{j+1})\mathcal{S}_{\mathcal{C}_2\Theta}(\vec{q}_{j})\\
& = \mathcal{S}^\dagger_{\mathcal{C}_2\Theta}(\vec{q}_{j+1}){\mathcal M} (\vec{q}_j,\vec{q}_{j+1})\mathcal{S}_{\mathcal{C}_2\Theta}(\vec{q}_{j})
\end{align*}
Next, we use that the overlap matrix between the parallel-transported states can 
be series expanded in the mesh-size $1/\mu$ as
\begin{align*}
{\mathcal M}(\vec{q}_j,\vec{q}_{j+1})&= 1 + \frac{1}{\mu}\mathcal{H}_1 + \mathcal{O}(1/M^2).
\end{align*}
Moreover, 
the sewing matrix
$\mathcal{S}_{\mathcal{C}_2\Theta}(\vec{q}_{j+1})\rightarrow\mathcal{S}_{\mathcal{C}_2\Theta}(\vec{q}_{j})$ in the 
$1 / \mu \rightarrow 0$ limit. 
Hence, we can also expand $\mathcal{S}_{\mathcal{C}_2\Theta}(\vec{q}_{j+1})$ 
in a series
\begin{align*}
\mathcal{S}_{\mathcal{C}_2\Theta}(\vec{q}_{j+1})& = \mathcal{S}_{\mathcal{C}_2\Theta}(\vec{q}_{j})\exp{(i \frac{1}{M}\mathcal{H}_2)}\\
&=\mathcal{S}_{\mathcal{C}_2\Theta}(\vec{q}_{j})(1+i \frac{1}{M}\mathcal{H}_2 + \mathcal{O}(1/M^2))
\end{align*}
Crucially, both $\mathcal{H}_1$ and $\mathcal{H}_2$ are Hermitian matrices. 
We can now rewrite the relation combining the overlap matrix with the ${\mathcal C}_2 \Theta$ sewing matrices written above, and ignoring terms $\propto \mu^{-2}$ we have
\begin{align*}
1 + \frac{1}{\mu}\mathcal{H}_1 & = (1-i \frac{1}{\mu}\mathcal{H}_2)\mathcal{S}^\dagger_{\mathcal{C}_2\Theta}(\vec{q}_{j})(1 + \frac{1}{\mu}\mathcal{H}_1)\mathcal{S}_{\mathcal{C}_2\Theta}(\vec{q}_{j})\\
&=1 -i \frac{1}{\mu}\mathcal{H}_2+ \frac{1}{\mu}\mathcal{S}^\dagger_{\mathcal{C}_2\Theta}(\vec{q}_{j})\mathcal{H}_1\mathcal{S}_{\mathcal{C}_2\Theta}(\vec{q}_{j}).
\end{align*}
In order for the right-hand side to be Hermitian, as the left-hand side, we find $\mathcal{H}_2 = 0$. Hence, we find that $\mathcal{S}_{\mathcal{C}_2\Theta}(\vec{q}_j) = \mathcal{S}_{\mathcal{C}_2\Theta}(\vec{q}_{j+1})$
, thus verifying that the parallel transported states fulfill our symmetric gauge requirement
along the line connecting $\Gamma$ and $Y$. 

\end{appendix}

\end{document}